\documentclass[a4paper,11pt,dvipsnames]{article}

\pdfoutput=1 %
\usepackage{jheppub} %
\usepackage{graphicx}
\usepackage{cancel}
\usepackage{color}
\usepackage{psfrag}
\usepackage{braket}
\usepackage{subfig}
\usepackage{amsfonts,amsmath,amssymb}
\usepackage{slashed}
\usepackage[usenames,dvipsnames]{xcolor}
\usepackage{url}
\usepackage{longtable}
\usepackage{comment}
\usepackage{rotating} 
\usepackage{blkarray}
\mathchardef\mhyphen="2D

\usepackage{multirow}
\usepackage{booktabs}

\usepackage{placeins}

\newcommand{\nc}{\newcommand}

\newcommand{ \mysmall}[1]{\scriptscriptstyle #1} 

\renewcommand{\bar}[1]{\overline{#1}}

\newcommand{\aem}{\alpha_{\text{em}}}
\newcommand{\be }{\begin{equation}}   \nc{\ee }{\end{equation}}
\newcommand{\bea}{\begin{eqnarray}}   \nc{\eea}{\end{eqnarray}}
\newcommand{\baa}{\begin{array}}      \nc{\eaa}{\end{array}}
\newcommand{\bit}{\begin{itemize}}    \nc{\eit}{\end{itemize}}
\newcommand{\ben}{\begin{enumerate}}  \nc{\een}{\end{enumerate}}

\newcommand{\Eq}[1]{Eq.~(\ref{#1})}

\newcommand{\RN}[1]{%
  \textup{\uppercase\expandafter{\romannumeral#1}}%
}

\newcommand{\fop}{\mathcal{F}}

\newcommand{\A}{{\cal A}}
\newcommand{\eq}[1]{\begin{equation} #1 \end{equation}}
\newcommand{\eqa}[1]{\begin{eqnarray} #1 \end{eqnarray}}
\renewcommand{\L}{{\cal L}}
\newcommand{\sss}{\scriptscriptstyle}
\newcommand{\C}{{\cal C}}
\renewcommand{\O}{{\cal O}}

\newcommand{\av}[1]{\langle #1 \rangle}

\begin{document}
\preprint{\vbox{\hbox{}}} 
\title{\boldmath $B$ physics Beyond the Standard Model at One Loop:\\
Complete Renormalization Group Evolution\\ below the Electroweak Scale}

\author{Jason Aebischer,} \emailAdd{aebischer@itp.unibe.ch}
\author{Matteo Fael,}\emailAdd{fael@itp.unibe.ch}
\author{Christoph Greub}\emailAdd{greub@itp.unibe.ch}
\author{and Javier Virto}\emailAdd{jvirto@mit.edu}

\affiliation{Albert Einstein Center for Fundamental Physics, Institute
  for Theoretical Physics,\\ University of Bern, CH-3012 Bern,
  Switzerland.}

\abstract{General analyses of $B$-physics processes beyond the Standard Model require accounting for
operator mixing in the renormalization-group evolution from the matching scale down to the typical
scale of $B$ physics. For this purpose the anomalous dimensions of the full set of
local dimension-six operators beyond the Standard Model are needed. We present here for the first time a complete and non-redundant set of dimension-six operators relevant for $B$-meson mixing and decay, 
together with the complete one-loop anomalous dimensions in QCD and QED.
These results are an important step towards the automation of general New Physics analyses.}
\maketitle
\flushbottom

\section{Introduction \label{sec:intro}}

\emph{$B$ physics} is the physics related to the decay and mixing of $B$ mesons. These processes require a change of
beauty ($B$) quantum number, $\Delta B\ne 0$,
and must therefore be mediated either by weak interactions or by physics beyond the Standard Model (SM).
Weak interactions (including interactions with the Higgs field) are mediated by heavy particles with masses of order of the
Electroweak (EW) scale, around $\mu_{\rm EW} \simeq 100$~GeV. This scale is very large in comparison to the center-of-mass energy of
$B$-physics processes, around $m_b \simeq 5$~GeV, and thus the weak
interaction can be regarded as a \emph{local interaction},
``factorizing'' from the non-perturbative physics of mesons and of the strong and electromagnetic
effects operating at these low energy scales.
If the physics beyond the Standard Model (BSM) is also mediated by new particles with masses much larger than the
$B$-physics scale, BSM interactions will be also approximately local. 
This will be an implicit assumption throughout the paper:
that the BSM scale $\Lambda$ is at least of the order of the EW scale, $\Lambda \gtrsim \mu_{\rm EW} \gg m_b$.
Therefore both in the case of Weak and BSM interactions, corrections beyond the leading local contribution are suppressed by additional powers of $m_b/\mu_{\rm EW}$ or $m_b/\Lambda$,
and completely negligible in comparison with current uncertainties in the computations of the leading matrix elements. As a result, $B$ physics within and beyond the SM is well described by an effective
Lagrangian which includes QCD and QED coupled to all six leptons and the five lightest quarks, plus a full set of \emph{local}
dimension-six operators consistent with the field content and gauge symmetry below the EW scale:
\eq{
\L_{\rm\sss WET} = \L_{\rm \sss QCD+QED}^{(u,d,c,s,b,e,\mu,\tau,\nu_e,\nu_\mu,\nu_\tau)} + \sum_i \C^{(6)}_i\,\O^{(6)}_i\ .
}
Here $\O_i^{(6)}$ denote the (bare) dimension-six local operators, and
$\C_i^{(6)}$ are the corresponding (bare) couplings or \emph{Wilson coefficients}.
This effective theory is called the ``Weak Effective Theory'' (WET).
For a pedagogical account of the standard formalism we refer to the classical reviews in Refs.~\cite{9512380,9806471}.

One of the convenient features of Effective Field Theory is the framework it provides for the resummation of large logarithms.
In $B$ physics, the perturbative hard-gluon corrections to physical amplitudes lead to expansions of the type
$\A = \sum_n a_n(m_b, \mu_{\rm EW}, \Lambda,\dots)  \alpha_s(m_b)^n$, where $a_n$ contain terms proportional to
$\log^n{(m_b/\mu_{\rm EW})}$. Thus the series expansion contains sub-series of the type
$\sum_n b_n [\alpha_s(m_b) \log{(m_b/\mu_{\rm EW})]^n}$. Since $\mu_{\sss\rm EW} \gg m_b$, the logarithm is large
and these sub-series do not have good convergence properties, and they must be resummed.
This resummation can be performed with Renormalization Group (RG) methods
within the WET, and leads to a reorganization of the perturbative series.
This requires to know the renormalization-scale dependence of the renormalized operators in the effective theory,
which is given by the \emph{anomalous dimensions}.

The WET has been studied extensively as an effective theory of the Standard Model below the EW scale.
Matching the SM to the WET perturbatively leads to initial conditions for the Wilson coefficients
$\C_i(\mu_0\sim \mu_{\rm\sss EW})$ as functions of the SM parameters (see e.g.~\cite{9910220}).
However, in the SM many of the matching conditions are negligible, and it is then conventional to restrict the
operator basis to a subset which is closed under renormalization and contains all the operators with non-negligible
matching conditions. This basis may be called the ``SM operator basis''. The anomalous dimensions of the SM operator
basis are known to high perturbative
orders~\cite{9910220,9211304,Ciuchini:1993vr,9409454,9612313,9711280,9711266,0306079,0411071,0504194,0612329}.

Beyond the SM it will typically be the case that operators outside the SM operator basis are generated with
relevant matching coefficients. This happens for example when matching the WET to a general set of dimension-six terms
in the SM~\cite{1008.4884,1512.02830,Aebischer:2016xmn}.
Thus, the BSM $B$-physics toolkit should contain the full set of anomalous dimensions,
at least to the leading non-trivial order. Many bits and pieces of the full anomalous dimension matrix (ADM) relevant for
BSM physics have been calculated in the past, but no complete account is available to date. It is the purpose of this paper
to collect and complete the calculation of the one-loop anomalous dimensions
in QCD and QED for the full operator basis in
$B$ physics.

We start in Section~\ref{set-up} defining the Weak Effective Theory beyond the Standard Model and constructing a complete and non-redundant operator basis.
In Section~\ref{renormalization} we outline the QCD and QED renormalization of the effective theory.
In Section~\ref{ADM} we discuss the calculation of the full set of one-loop anomalous dimensions and collect the results. 
In Section~\ref{RGE} we solve the Renormalization Group Equation by constructing the evolution matrix and discuss the one-loop QCD and QED scale dependence of the
Wilson coefficients. A brief numerical discussion is presented in Section~\ref{sec:spectra}.
In Section~\ref{conclusion} we conclude with a summary.
The appendices contain: A description of the complete set of results in electronic format attached to this paper
(App.~\ref{notebook}), the Fierz identities needed to make the operator basis minimal (App.~\ref{fierz}),
and the procedure to translate our results to other more traditional bases used in the literature, first for
magnetic and semileptonic operators (App.~\ref{SemileptonicBasis}), and then for 4-quark operators, together with a careful comparison of
different sets of results with previous calculations (App.~\ref{changebasis}).

\subsection*{Conventions}

Throughout the paper we use the following conventions and definitions: we use the convention
$\sigma_{\mu\nu} \equiv \frac{i}2 [\gamma_\mu,\gamma_\nu]$, and define the strings of gamma matrices
\eq{
\gamma_{\mu\nu\rho}\equiv \gamma_\mu \gamma_\nu \gamma_\rho\ ,\quad
\gamma_{\mu\nu\rho\sigma}\equiv \gamma_\mu \gamma_\nu \gamma_\rho \gamma_\sigma \ .
\notag
}
The Dirac left- and right-handed projectors are defined as $P_L \equiv (1 - \gamma_5)/2$ and $P_R \equiv (1 + \gamma_5)/2$, with the 4-dimensional $\gamma_5$ defined
as $\gamma_5\equiv -\frac{i}{4!} \epsilon^{\mu\nu\rho\sigma} \gamma_{\mu\nu\rho\sigma}$. With this definition, the following relations hold in $D=4$:
\eqa{
\gamma_{\mu\nu\rho} &=& g_{\mu\nu} \gamma_\rho - g_{\mu\rho} \gamma_\nu + g_{\nu\rho} \gamma_\mu
+ i \epsilon_{\mu\nu\rho\alpha}\, \gamma^\alpha \gamma_5 \label{eqn:idthreegammas},\\[2mm]
\gamma_{\mu\nu\rho\sigma} &=&
g_{\mu\sigma} g_{\nu\rho} - g_{\mu\rho} g_{\nu\sigma} + g_{\mu\nu} g_{\rho\sigma}
+ i\big[
g_{\nu\sigma}  \sigma_{\mu\rho} + \sigma_{\nu\sigma}  g_{\mu\rho} - g_{\rho\sigma} \sigma_{\mu\nu}
- \sigma_{\mu\sigma}  g_{\nu\rho}  \notag \\[1mm]
&&
- g_{\mu\sigma}  \sigma_{\nu\rho} - \sigma_{\rho\sigma}  g_{\mu\nu}
\big] - i  \epsilon_{\mu\nu\rho\sigma}\, \gamma_5, \label{eqn:idfourgammas}\\[2mm]
\sigma_{\mu\nu} \gamma_5 &=& \frac{i}2 \epsilon_{\mu\nu\alpha\beta}\ \sigma^{\alpha\beta} . \label{eqn:idsigma}
}
The totally antisymmetric tensor is defined such that $\epsilon^{0123} = - \epsilon_{0123} = +1$.
Throughout this paper we will use naive dimensional regularization with anticommuting
$\gamma_5$. This is convenient since our choice of basis will ensure that no
Dirac traces with $\gamma_5$ have to be evaluated~\cite{9711280}.
For conjugate fields we use the notation $\psi^c = C \bar \psi^{\, \sss T}$ and $\bar \psi^{\, c} \equiv \psi^{\sss T} C$, where $C$ denotes the charge-conjugation matrix. Useful relations are: $C^T = C^\dagger = C^{-1} = -C$
and $C\,\Gamma\,C^T = \eta_\Gamma\,\Gamma$, with $\eta_{1,\gamma_5,\gamma_\mu \gamma_5} = 1$ and
$\eta_{\gamma_\mu, \sigma_{\mu\nu}}=-1$~\cite{Denner:1992vza}.

Finally, our convention for QED and QCD covariant derivatives is such that
\eq{
{\cal D}_\mu \,\psi = (\partial_\mu + i\, e\,Q_\psi {\cal A}_\mu + i g_s T^a {\cal G}_\mu^a ) \,\psi
}
with $Q_e = -1$. 
The field-strength tensors are then defined by $ie\,Q_\psi F_{\mu\nu} + i g_s T^a G^a_{\mu\nu} = [{\cal D}_\mu,{\cal D}_\nu]$.

\begin{table}
\renewcommand*{\arraystretch}{1.00}
\begin{center}
\begin{tabular}{@{}lccccl@{}}
\toprule[1.4pt] 
Class  &  Flavour structure & Number of Ops.  & Other flavours    & \hspace{2mm} ADM \hspace{2mm} & Example process\\
\midrule[1.4pt]
Class I & $\bar s b\,\bar s b$ & 5+3  &  $\bar d b\,\bar d b$   & $\hat \gamma_{\rm\sss I}$ & $B_q - \bar B_q$ mixing\\
\midrule
Class II & $\bar u b\,\bar \ell \nu_{\ell'}$ & $(2+3) \times 9$  &  $\bar c b\,\bar \ell \nu_{\ell'}$   & $\hat \gamma_{\rm\sss II}$ 
& $\bar B_d \to \pi^+ \mu^- \bar \nu$\\
\midrule
\multirow{3}{*}{Class III} & \multirow{3}{*}{$\bar s b\,\bar u c$} & \multirow{3}{*}{10+10} & $\bar s b\,\bar c u$  & \multirow{3}{*}{$\hat \gamma_{\rm\sss III}$}
& \multirow{3}{*}{$B^- \to \bar D^0 K^-$}\\
&&& $\bar d b\,\bar u c$ &&  \\
&&& $\bar d b\,\bar c u$ && \\
\midrule
\multirow{2}{*}{Class IV} & \multirow{2}{*}{$\bar s b\,\bar s d$} & \multirow{2}{*}{5+5} & $\bar d b\,\bar d s$  & \multirow{2}{*}{$\hat \gamma_{\rm\sss IV}$}  & \multirow{2}{*}{$B^- \to \bar K^0 K^-$} \\
&&&  $\bar b s\,\bar b d$ && \\
\midrule
\multirow{3}{*}{Class V} & $\bar s b\,\bar q q$ & \multirow{3}{*}{57+57} & $\bar d b\,\bar q q$
& \multirow{3}{*}{$\hat \gamma_{\rm\sss V}$} & $\bar B_d \to D^+ D_s^-$ \\
&  $\bar s b\,F$, $\bar s b\,G$ && $\bar d b\,F$, $\bar d b\,G$ && $\bar B_d \to X_s \gamma$ \\
&  $\bar s b\,\bar \ell \ell$ && $\bar d b\,\bar \ell \ell$ && $B^- \to K^- \mu^+\mu^-$ \\
\midrule
Class Vb & $\bar s b\,\bar \ell \ell'$, $\ell\ne \ell'$ & $(5+5)\times 6$ & $\bar d b\,\bar \ell \ell'$  & $\hat \gamma_{\rm\sss Vb}$
& $\bar B_s \to \mu^- \tau^+$ \\
\midrule
Class V$\nu$ & $\bar s b\,\bar \nu_\ell \nu_{\ell'}$ & $(1+1)\times 9$ & $\bar d b\,\bar \nu_\ell \nu_{\ell'}$  & zero
& $B^- \to K^- \bar \nu \nu$ \\
\midrule
%
\multirow{3}{*}{Class VI} & $\bar u b\,\bar \ell  \nu^c $  & $(3+2)\times 9$ & $\bar c b\,\bar \ell \nu^c $ & $\hat \gamma_{\rm\sss VIIa}$
& $\bar B_d \to \pi^+ \mu^- \nu $ \\
\cmidrule{2-6}
& $\bar s b\,\bar \nu \nu^c$  & $(2+1)\times 6$ & $\bar d b\,\bar \nu \nu^c$ & $\hat \gamma_{\rm\sss VIb}$
& $B^- \to K^- \bar \nu \bar \nu$ \\
\cmidrule{2-6}
& $\bar s b\, \bar \nu^c \nu$  & $(1+2)\times 6$ & $\bar d b\, \bar \nu^c \nu$ & $\hat \gamma_{\rm\sss VIc}$
& $B^- \to K^- \nu \nu$ \\
\midrule
\multirow{13}{*}{Class VII} & $\bar \ell^c b\,\bar c^c u$  & $(5+5)\times 3$ & None & $\hat \gamma_{\rm\sss VIIa}$
& $\bar B_d \to \Lambda_c^- e^+$ \\
\cmidrule{2-6}
& $\bar \ell^c b\,\bar u^c u$  & $(2+2)\times 3$ & $\bar \ell^c b\,\bar c^c c$ & $\hat \gamma_{\rm\sss VIIb}$
& $B_d \to p\ e^-$ \\
\cmidrule{2-6}
& \multirow{3}{*}{$\bar \nu^c b\,\bar u^c s$}  & \multirow{3}{*}{$5\times 3$} & $ \bar \nu^c b\,\bar u^c d$ & \multirow{3}{*}{$\hat \gamma_{\rm\sss VIIc}$}
& \multirow{3}{*}{$B^+ \to p\, \nu$} \\
&                                            &                    & $\bar\nu^c  b\,\bar c^c s$ &  \\
&                                            &                    & $\bar\nu^c  b\,\bar c^c d$ &  \\
\cmidrule{2-6}
& $\bar \nu^c b\,\bar u^c b$  & $2\times 3$ & $ \bar \nu^c b\,\bar c^c b$ & $\hat \gamma_{\rm\sss VIId}$ 
& $\Lambda_b^0 \to B_d\,\bar \nu$ \\
\cmidrule{2-6}
& $\bar \ell b\,\bar s^c d$  & $(5+5)\times 3$ & None & $\hat \gamma_{\rm\sss VIIe}$
& $\Lambda_b^0 \to K^+ e^-$ \\
\cmidrule{2-6}
& $\bar \ell^c b\,\bar s^c s$  & $(2+2)\times 3$ & $\bar \ell^c b\,\bar d^c d$ & $\hat \gamma_{\rm\sss VIIf}$
& $\Lambda_b^0 \to \pi^+ e^-$ \\
\cmidrule{2-6}
& $\bar \ell^c b\,\bar s^c b$  & $(2+2)\times 3$ & $\bar \ell^c b\,\bar d^c b$ & $\hat \gamma_{\rm\sss VIIg}$
& $\Xi_b^0 \to B^+ e^-$ \\
\cmidrule{2-6}
& \multirow{3}{*}{$\bar \nu b\,\bar u^c s$}  & \multirow{3}{*}{$5\times 3$} & $\bar \nu b\,\bar u^c d$ & \multirow{3}{*}{$\hat \gamma_{\rm\sss VIIh}$}
& \multirow{3}{*}{$B^+ \to p\, \bar \nu$} \\
&                                            &                    & $\bar \nu  b\,\bar c^c s$ &  \\
&                                            &                    & $\bar \nu  b\,\bar c^c d$ &  \\
\cmidrule{2-6}
& $\bar \nu b\,\bar u^c b$  & $2\times 3$ & $ \bar \nu b\,\bar c^c b$ & $\hat \gamma_{\rm\sss VIIi}$ 
& $\Lambda_b^0 \to B_d\, \nu$ \\
\bottomrule[1.4pt]
\end{tabular}
\end{center}
\vspace{-5mm}
\caption{\small  Summary list of non-redundant operators. The number of operators in each class is indicated by $(n+n')\times n_\ell$,
where $n$ is the number of different operators modulo lepton flavours, $n'$ is the number of operators with opposite chirality,
and $n_\ell$ accounts for the different leptonic flavours. All the operators with flavour structure given in the second column are
defined in Section~\ref{set-up}, while the ones in the fourth column are obtained by obvious replacements. The last column
lists examples of processes to which the corresponding class of operators contributes.}
\label{TableList}
\end{table}

\FloatBarrier

\newpage

\section{Complete Operator Basis Beyond the SM}
\label{set-up}

In this paper we consider a complete and non-redundant basis for $\Delta B\ne 0$ operators beyond the
Standard Model. However, these operators will not always correspond one-to-one to the operators traditionally chosen in the
SM operator basis, for which matching conditions and anomalous dimensions are very well known and standard.
In order to be able to use, on the one hand, these well-known SM results directly, and other hand, our results for
BSM operators, it is convenient to  separate SM and BSM contributions at the level of the Lagrangian:
\eq{
\L_{\rm\sss WET} = \L_{\rm \sss QCD+QED}^{(u,d,c,s,b,e,\mu,\tau,\nu_e,\nu_\mu,\nu_\tau)} 
+ \L_{\rm\sss EW}^{(6)}
+ \L_{\rm\sss BSM}^{(6)}\ .
\label{LWET}
}
Here $\L_{\rm\sss EW}^{(6)}$ and $\L_{\rm\sss BSM}^{(6)}$ are the effective Lagrangians resulting after integrating out the SM and the BSM heavy degrees of freedom, respectively.
The effective Lagrangian originating from BSM physics is
\begin{equation}
  \L_{\rm\sss BSM}^{(6)} =
 \frac{4 G_F}{\sqrt2} \sum_i \Big[ \ \C_i^{(0)}\,\O_i^{(0)} + h.c.\ \Big]\ ,
 \label{eqn:LNP}
\end{equation}
where the sum over $i$ runs over all the operator indices that will appear below.
The superindex~$(0)$ indicates that the Wilson coefficients and the operators in \Eq{eqn:LNP} are \emph{bare} quantities and must be renormalized.
The relationship between bare and renormalized quantities will be discussed in Section~\ref{renormalization}.
The coefficients $\C_i$ contain all BSM effects but no pure-SM ones. 
Thus the SM matching conditions determine $\L_{\rm\sss EW}^{(6)}$, and matching conditions involving BSM particles determine the $\C_i$. 
We organize the operators such that $\O_i$ and $\O_i^\dagger$ have $\Delta B >0$ and $\Delta B <0$, respectively.
They can be grouped into classes according to their flavour quantum numbers. 
This is useful because the flavour symmetries of QCD and QED imply that the different groups cannot mix into each other. 
A summary of the full list of non-redundant operators classified according to their flavour structure is given in Table~\ref{TableList}. 
In order to keep a unified notation for all classes, we introduce a generic
basis for four-fermion operators, after which we
list the operators in each class.

\subsection*{Generic Basis for Four-Fermion Operators}

We adopt (except for Class I) the following generic basis of four-fermion operators \emph{\`{a} la Chetyrkin-Misiak-Munz}~\cite{9711280}:
\begin{align}
\O_1 &= (\bar{\psi}_1 \,P_R\, \gamma_\mu\, \psi_2)\;   ( \bar{\psi}_3 \gamma^\mu \psi_4)\,, &
\O_2 &=(\bar{\psi}_1\,P_R\,  \gamma_\mu \,T^{\mysmall A}\psi_2)\;  
       (\bar{\psi}_3 \gamma^\mu\, T^{\mysmall A} \psi_4)\,, \notag\\[1mm]
\O_3&=(\bar{\psi}_1 \,P_R\, \gamma_{\mu\nu\rho}\, \psi_2)\; (\bar{\psi}_3 \gamma^{\mu\nu\rho} \psi_4)\,, &
\O_4&=(\bar{\psi}_1\,P_R\, \gamma_{\mu\nu\rho} T^{\mysmall A}\psi_2)\; 
     (\bar{\psi}_3 \gamma^{\mu\nu\rho}\, T^{\mysmall A} \psi_4)\,, \notag\\[1mm]
\O_5 &= (\bar{\psi}_1 \,P_R\, \psi_2) (\bar{\psi}_3\, \psi_4)\,, &
\O_6 &=(\bar{\psi}_1 \,P_R\, T^{\mysmall A} \psi_2) (\bar{\psi}_3\, T^{\mysmall A}\psi_4)\; , \label{eqn:CMMbasis}\\[1mm]
\O_7 &=(\bar{\psi}_1\, \,P_R\, \sigma^{\mu\nu} \, \psi_2) (\bar{\psi}_3 \,\sigma_{\mu\nu} \,\psi_4)\,, &
\O_8 &=(\bar{\psi}_1 \,P_R\, \sigma^{\mu\nu} \,T^{\mysmall A} \psi_2) (\bar{\psi}_3 \sigma_{\mu\nu}\,T^{\mysmall A} \psi_4)\,,\notag\\[1mm]
\O_9 &=(\bar{\psi}_1 \,P_R\, \gamma_{\mu\nu\rho\sigma} \,  \psi_2)\; (\bar{\psi}_3 \gamma^{\mu\nu\rho\sigma} \psi_4)\,, &
\O_{10} &=(\bar{\psi}_1 \,P_R\, \gamma_{\mu\nu\rho\sigma} \, T^{\mysmall A} \psi_2)\; 
          (\bar{\psi}_3 \gamma^{\mu\nu\rho\sigma}\, T^{\mysmall A}  \psi_4)\, , \notag
\end{align}
where $T^A$ are the $SU(3)$ generators in the fundamental representation.
In addition, operators with primed indices, obtained by interchanging $P_R \leftrightarrow P_L$, must be considered as well. These will also be referred to as operators with ``opposite chirality''.

The basis~\eqref{eqn:CMMbasis} has been used extensively used in higher-order calculations~\cite{9910220,
9612313,9711280,9711266,0306079,0411071,0504194,0612329} because it allows to avoid the evaluation of Dirac traces containing $\gamma_5$ to any order in QCD and QED, provided that $\psi_1$ and $\psi_2$ have different flavours.
In $D=4$, it can be rewritten in terms of the chiral basis --- with Dirac structure of the form $P_L \otimes P_L, P_L \otimes P_R, \gamma^\mu P_L \otimes \gamma_\mu P_L,\gamma^\mu P_L \otimes \gamma_\mu P_R, \sigma^{\mu\nu} P_L \otimes \sigma_{\mu\nu} P_L, \dots$  --- by means of the identities~(\ref{eqn:idthreegammas}-\ref{eqn:idsigma}), see App.~\ref{fierz}. 

The operators with an even index, $\O_{2,4,6,8,10}$, must be considered only for four-quark operators.
In addition, when $\psi_1 =\psi_3$ or $\psi_2 = \psi_4$ not all operators are independent and the basis can be reduced by means of Fierz identities. In such case we always choose to remove the operators with an even index (see Appendix~\ref{fierz} for more details).

\subsection*{Class I : $\mathbf{|\Delta B| = 2}$ operators}

For $|\Delta B| =2 $ operators we use the traditional ``SUSY basis''~\cite{Gabbiani:1996hi,Virto:2009wm} (but paying attention to the different
normalization in \Eq{eqn:LNP}). In the case of $|\Delta S|=2$ this basis
is given by 
\begin{align}
\O_1^{sbsb}   \,= & \, ({\bar s} \gamma_{\mu} P_{L} b)\, ({\bar s} \gamma^{\mu} P_{L} b)\,, &
\O_5^{sbsb}  \,= & \,({\bar s}_{\alpha} P_{L} b_{\beta})\, ({\bar s}_{\beta}  P_{R} b_{\alpha}) \,,   \notag \\[1mm]
\O_2^{sbsb}  \,= & \, ({\bar s}P_{L} b)\, ({\bar s}  P_{L} b)  \,,  &
\O_{1'}^{sbsb}   \,= & \, ({\bar s} \gamma_{\mu} P_{R} b)\, ({\bar s} \gamma^{\mu} P_{R} b)\,, \notag \\[1mm]
\O_3^{sbsb}  \,= &\, ({\bar s}_{\alpha} P_{L} b_{\beta})\, ({\bar s}_{\beta}  P_{L} b_{\alpha}) \,,  &    
\O_{2'}^{sbsb}  \,= & \, ({\bar s}P_{R} b)\, ({\bar s}  P_{R} b)  \,, \notag \\[1mm]
\O_4^{sbsb}   \,= & \, ({\bar s} P_{L} b)\, ({\bar s}  P_{R} b)  \,,  &
\O_{3'}^{sbsb}  \,= &\, ({\bar s}_{\alpha} P_{R} b_{\beta})\, ({\bar s}_{\beta}  P_{R} b_{\alpha})   \,,
&
\label{opbasisDF2}
\end{align}
where we denote with primed indices the operators with opposite chirality. The corresponding $|\Delta S|=0$
operators $\O_i^{dbdb}$ are obtained from $\O_i^{sbsb}$ by performing the substitution $s\to d$.

\subsection*{Class II : $\mathbf{|\Delta B| = 1}$ semileptonic operators}

In semileptonic operators we allow for lepton-flavour violation and non-universality, with $\ell,\ell^{\prime}\in\{e,\mu,\tau\}$. 
Neutrinos are assumed to be left-handed and we shortly denote them by $\nu_\ell \equiv \nu_{\ell, \, L}$.
The operators can be either $|\Delta I|=1/2$ or $|\Delta C|=1$; the basis in the former case is given by
\begin{align}
\O_{1}^{ub\ell\ell'} &= \left( \bar{u} \,P_R\, \gamma^\mu \, b \right)  \left( \bar{\ell} \, \gamma_\mu \,\nu_{\ell'} \right)\,, &
\O_{5}^{ub\ell\ell'} &= \left( \bar{u}\, P_R\, b \right)  \left( \bar{\ell}  \,\nu_{\ell'} \right)\,, &
\\[1mm]
\O_{1'}^{ub\ell\ell'} &= \left( \bar{u} \, P_L\, \gamma^\mu \, b \right)  \left( \bar{\ell} \, \gamma_\mu \,\nu_{\ell'} \right)\,, &
\O_{5'}^{ub\ell\ell'} &= \left( \bar{u}\, P_L \,b \right)  \left( \bar{\ell} \, \nu_{\ell'} \right)\,, &
\O_{7'}^{ub\ell\ell'} &= \left( \bar{u} \, P_L\, \sigma^{\mu\nu}\,b \right)  \left( \bar{\ell} \, \sigma_{\mu\nu} \,\nu_{\ell'} \right)\,.
\notag
\end{align}
The corresponding $|\Delta C| =1$ operators $\O_i^{cb\ell\ell'}$ are obtained from $\O_i^{ub\ell\ell'}$ by performing the substitution $u\to c$.

\subsection*{Class III : $\mathbf{|\Delta B| = |\Delta C| =1}$  four-quark operators}

A complete basis for $|\Delta B| = |\Delta C| = |\Delta S| = 1$ operators is given by
\begin{align}
\O_1^{sbuc} &= (\bar{s} \,P_R\, \gamma_\mu\, b)\;   ( \bar{u} \gamma^\mu c)\,, &
\O_2^{sbuc} &=(\bar{s}\,P_R\,  \gamma_\mu \,T^{\mysmall A}b)\;  (\bar{u} \gamma^\mu\, T^{\mysmall A} c)\,, \notag\\[1mm]
\O_3^{sbuc}&=(\bar{s} \,P_R\, \gamma_{\mu\nu\rho}\, b)\; (\bar{u} \gamma^{\mu\nu\rho} c)\,, &
\O_4^{sbuc} &=(\bar{s}\,P_R\, \gamma_{\mu\nu\rho} T^{\mysmall A}b)\; (\bar{u} \gamma^{\mu\nu\rho}\, T^{\mysmall A} c)\,, \notag\\[1mm]
\O_5^{sbuc} &= (\bar{s} \,P_R\, b) (\bar{u}\, c)\,, &
\O_6^{sbuc} &=(\bar{s} \,P_R\, T^{\mysmall A} b) (\bar{u}\, T^{\mysmall A}c)\; , \\[1mm]
\O_7^{sbuc}&=(\bar{s}\, \,P_R\, \sigma^{\mu\nu} \, b) (\bar{u} \,\sigma_{\mu\nu} \,c)\,, &
\O_8^{sbuc} &=(\bar{s} \,P_R\, \sigma^{\mu\nu} \,T^{\mysmall A} b) (\bar{u} \sigma_{\mu\nu}\,T^{\mysmall A} c)\,,\notag\\[1mm]
\O_9^{sbuc}&=(\bar{s} \,P_R\, \gamma_{\mu\nu\rho\sigma} \,  b)\; (\bar{u} \gamma^{\mu\nu\rho\sigma} c)\,, &
\O_{10}^{sbuc} &=(\bar{s} \,P_R\, \gamma_{\mu\nu\rho\sigma} \, T^{\mysmall A} b)\; (\bar{u} \gamma^{\mu\nu\rho\sigma}\, T^{\mysmall A}  c)\,, \notag
\end{align}
plus the analogous set with opposite chirality
\eq{
\O_{i'}^{sbuc} = \O_{i}^{sbuc} \Big|_{P_{L,R}\to P_{R,L}}\ ,
}
and similarly for the operators $\O_i^{sbcu}$ with opposite sign for $\Delta C$.
The corresponding operators $\O_i^{dbuc}$ and $\O_i^{dbcu}$ with $|\Delta S| =0$ ($|\Delta I|=1$) are obtained
from $\O_i^{sbuc}$ and $\O_i^{sbcu}$ by performing the substitution $s\to d$.

\subsection*{Class IV : $\mathbf{|\Delta B| =1, \, |\Delta S| = 2}$  four-quark operators}

A complete basis for $|\Delta B| =1, |\Delta S| = 2$ operators is given by
\begin{align}
\O_1^{sbsd} &= (\bar{s}\, \gamma_\mu\,P_L \, b)\;   ( \bar{s} \gamma^\mu \,d )\,, &
\O_{1'}^{sbsd} &=(\bar{s}\, \gamma_\mu\,P_R \,b)\; (\bar{s} \gamma^\mu\, d )\,, \notag\\[1mm]
\O_3^{sbsd}&=(\bar{s}\,\gamma_{\mu\nu\rho}\,P_L \, b)\;( \bar{s} \gamma^{\mu\nu\rho} d)\,, &
\O_{3'}^{sbsd} &=(\bar{s}\,\, \gamma_{\mu\nu\rho}\,P_R \, b)\;(\bar{s} \gamma^{\mu\nu\rho}\, d)\,, \notag\\[1mm]
\O_5^{sbsd} &= (\bar{s}\,P_L\, b) (\bar{s}\, d)\,, &
\O_{5'}^{sbsd} &=(\bar{s}\,P_R\, b) (\bar{s}\, d)\; ,  \notag\\[1mm]
\O_7^{sbsd}&=(\bar{s}\, \sigma^{\mu\nu}\,P_L\, b) (\bar{s} \,\sigma_{\mu\nu} \,d)\,, &
\O_{7'}^{sbsd} &=(\bar{s}\,\sigma^{\mu\nu}\,P_R\, b) (\bar{s} \,\sigma_{\mu\nu} \,d)\,,\label{sbsd}\\[1mm]
\O_9^{sbsd}&=(\bar{s}\,\gamma_{\mu\nu\rho\sigma}\,P_L \, b)\;( \bar{s} \gamma^{\mu\nu\rho\sigma} d)\,, &
\O_{9'}^{sbsd} &=(\bar{s}\,\gamma_{\mu\nu\rho\sigma}\,P_R \, b)\;( \bar{s} \gamma^{\mu\nu\rho\sigma} d)\,. \notag
\end{align}
Here we have chosen a different basis compared to the generic basis of Eq.~(\ref{eqn:CMMbasis}) to avoid mixing between primed and non-primed operators.
All color-octet operators are Fierz-redundant and have been omitted (see App.~\ref{fierz}).
The corresponding set of  $|\Delta B| = |\Delta I| =|\Delta S|=1$
operators $\O_i^{dbds}$ are obtained from $\O_i^{sbsd}$ by performing the substitution $s \leftrightarrow d$.
For completeness, the $|\Delta B| = 2$ operators $\O_i^{bsbd}$ are also included in Table~\ref{TableList}.

\subsection*{Class V : $\mathbf{|\Delta B| = 1,\ |\Delta C| = 0}$  operators}

There are three classes of such operators: Magnetic, hadronic (four-quark) and semileptonic operators.
In the case of $|\Delta S| = 1$, these are chosen as:

\noindent $\blacktriangleright$ Magnetic penguins:
\begin{align}
\O_{7\gamma}^{s} &= \frac{e}{g_s^2} m_b \,  ( \bar{s}  \, P_R\, \sigma_{\mu\nu}  \,  b ) \; F^{\mu\nu},&
\O_{8g}^s &= \frac{1}{g_s} m_b \, (\bar{s} \, P_R\, \sigma_{\mu\nu} \,\, T^{ A} \, b ) \; G^{\mu\nu}_A\,,\notag\\[1mm]
\O_{7'\gamma}^{s} &= \frac{e}{g_s^2} m_b \,  ( \bar{s}  \, P_L\, \sigma_{\mu\nu} \,  b ) \; F^{\mu\nu},&
\O_{8'g}^s &= \frac{1}{g_s} m_b \, (\bar{s} \, P_L\, \sigma_{\mu\nu} \,\, T^{ A} \, b ) \; G^{\mu\nu}_A\,, \label{sbM}
\end{align}
Our conventions for the field-strength tensors have been specified in the previous section.

\noindent $\blacktriangleright$ Four-quark ($q \neq s$):
\begin{align}
\O_1^{sbqq} &= (\bar{s}\,P_R\, \gamma_\mu  b)\;   ( \bar{q} \gamma^\mu q )\,, &
\O_2^{sbqq} &=(\bar{s}\,P_R\,  \gamma_\mu \,T^{\mysmall A}b)\; (\bar{q} \gamma^\mu\, T^{\mysmall A} q )\,, \notag\\[1mm]
\O_3^{sbqq}&=(\bar{s}\,P_R\,\gamma_{\mu\nu\rho} \, b)\;( \bar{q} \gamma^{\mu\nu\rho} q)\,, &
\O_4^{sbqq} &=(\bar{s}\,P_R\, \gamma_{\mu\nu\rho}  T^{\mysmall A}b)\;(\bar{q} \gamma^{\mu\nu\rho}\, T^{\mysmall A} q)\,, \notag\\[1mm]
\O_5^{sbqq} &= (\bar{s}\,P_R\, b) (\bar{q}\, q)\,, &
\O_6^{sbqq} &=(\bar{s}\,P_R\,T^{\mysmall A} b) (\bar{q}\, T^{\mysmall A}q)\; ,  \notag\\[1mm]
\O_7^{sbqq}&=(\bar{s}\,P_R\, \sigma^{\mu\nu} \, b) (\bar{q} \,\sigma_{\mu\nu} \,q)\,, &
\O_8^{sbqq} &=(\bar{s}\,P_R\,\sigma^{\mu\nu} \,T^{\mysmall A} b) (\bar{q} \sigma_{\mu\nu}\,T^{\mysmall A} q)\,,\label{sbqq}\\[1mm]
\O_9^{sbqq}&=(\bar{s}\,P_R\,\gamma_{\mu\nu\rho\sigma}\,b)\;( \bar{q} \gamma^{\mu\nu\rho\sigma} q)\,, &
\O_{10}^{sbqq} &=(\bar{s}\,P_R\,\gamma_{\mu\nu\rho\sigma} \, T^{\mysmall A} b)\;(\bar{q} \gamma^{\mu\nu\rho\sigma}\, T^{\mysmall A}  q)\,, \notag
\end{align}
where $q=\{ u, d, c\}$. In the case of $q= b$, the color-octet operators $\O^{sbbb}_{2,4,6,8,10}$ are Fierz-equivalent
to the color-singlet ones (see App.~\ref{fierz} for details) and are not included in the basis.
In addition, (for $q=\{ u, d, c, b\}$) the analogous set with opposite chirality is needed:
\eq{
\O_{i'}^{sbqq} = \O_{i}^{sbqq} \Big|_{P_{L,R}\to P_{R,L}}
}
The case $q=s$ needs a separate discussion because it is convenient to group primed and unprimed operators in a different
manner, which simplifies the mixing pattern:

\noindent $\blacktriangleright$ Four-quark ($q = s$):
\begin{align}
\O_1^{sbss} &= (\bar{s}\, \gamma_\mu\,P_L \, b)\;   ( \bar{s} \gamma^\mu \,s )\,, &
\O_{1'}^{sbss} &=(\bar{s}\, \gamma_\mu\,P_R \,b)\; (\bar{s} \gamma^\mu\, s )\,, \notag\\[1mm]
\O_3^{sbss}&=(\bar{s}\,\gamma_{\mu\nu\rho}\,P_L \, b)\;( \bar{s} \gamma^{\mu\nu\rho} s)\,, &
\O_{3'}^{sbss} &=(\bar{s}\,\, \gamma_{\mu\nu\rho}\,P_R \, b)\;(\bar{s} \gamma^{\mu\nu\rho}\, s)\,, \notag\\[1mm]
\O_5^{sbss} &= (\bar{s}\,P_L\, b) (\bar{s}\, s)\,, &
\O_{5'}^{sbss} &=(\bar{s}\,P_R\, b) (\bar{s}\, s)\; ,  \notag\\[1mm]
\O_7^{sbss}&=(\bar{s}\, \sigma^{\mu\nu}\,P_L\, b) (\bar{s} \,\sigma_{\mu\nu} \,s)\,, &
\O_{7'}^{sbss} &=(\bar{s}\,\sigma^{\mu\nu}\,P_R\, b) (\bar{s} \,\sigma_{\mu\nu} \,s)\,,\label{sbss}\\[1mm]
\O_9^{sbss}&=(\bar{s}\,\gamma_{\mu\nu\rho\sigma}\,P_L \, b)\;( \bar{s} \gamma^{\mu\nu\rho\sigma} s)\,, &
\O_{9'}^{sbss} &=(\bar{s}\,\gamma_{\mu\nu\rho\sigma}\,P_R \, b)\;( \bar{s} \gamma^{\mu\nu\rho\sigma} s)\,. \notag
\end{align}
Again, the color-octet operators are Fierz-redundant and have been omitted (see App.~\ref{fierz}).

\noindent $\blacktriangleright$ Semileptonic :
\begin{align}
\O_1^{sb\ell\ell'} &= (\bar{s}\, P_R\, \gamma_\mu \, b)\;   ( \bar{\ell} \gamma^\mu \ell' )\,, &
\O_{1'}^{sb\ell\ell'} &=(\bar{s}\, P_L \gamma_\mu \,b)\; (\bar{\ell} \gamma^\mu\, \ell' )\,, \notag\\[1mm]
\O_3^{sb\ell\ell'}&=(\bar{s}\,P_R\,\gamma_{\mu\nu\rho}  \,b)\;( \bar{\ell} \gamma^{\mu\nu\rho} \ell')\,, &
\O_{3'}^{sb\ell\ell'} &=(\bar{s}\,P_L\, \gamma_{\mu\nu\rho}  \,b)\;(\bar{\ell} \gamma^{\mu\nu\rho}\, \ell')\,, \notag\\[1mm]
\O_5^{sb\ell\ell'} &= (\bar{s}\,P_R \,b) (\bar{\ell}\, \ell')\,, &
\O_{5'}^{sb\ell\ell'} &=(\bar{s}\,P_L \,b) (\bar{\ell}\, \ell')\; ,  \notag\\[1mm]
\O_7^{sb\ell\ell'}&=(\bar{s}\, P_R\, \sigma^{\mu\nu} \,b) (\bar{\ell} \,\sigma_{\mu\nu} \,\ell')\,, &
\O_{7'}^{sb\ell\ell'} &=(\bar{s}\, P_L\,\sigma^{\mu\nu}  \,b) (\bar{\ell} \,\sigma_{\mu\nu} \,\ell')\,, \label{sbll}\\[1mm]
\O_9^{sb\ell\ell'}&=(\bar{s}\, P_R\,\gamma_{\mu\nu\rho\sigma}  \,b)\;( \bar{\ell} \gamma^{\mu\nu\rho\sigma} \ell')\,, &
\O_{9'}^{sb\ell\ell'} &=(\bar{s}\,P_L\,\gamma_{\mu\nu\rho\sigma}  \,b)\;( \bar{\ell} \gamma^{\mu\nu\rho\sigma} \ell')\,, \notag\\[3mm]
\O_{\nu \,1}^{sb\ell\ell'} &= (\bar{s}\,P_R\, \gamma_\mu \,b)\;   ( \bar{\nu}_\ell \gamma^\mu \nu_{\ell'} )\,, &
\O_{\nu \,1'}^{sb\ell\ell'}&=(\bar{s}\, P_L\, \gamma_\mu  \,b)\; (\bar{\nu}_\ell \gamma^\mu \nu_{\ell'} )\,. 
\label{sbnunu}
\end{align}
In semileptonic operators we also allow for lepton-flavour non-universality, and lepton-flavour violation. The later
case ($\ell\ne\ell'$) is referred to as {\bf Class~Vb}, while the case with two neutrinos is referred to as
{\bf Class~V$\nu$}.

The corresponding $|\Delta S| = 0$ ($|\Delta I| = 1/2$) operators are obtained from Eqs.~(\ref{sbM})-(\ref{sbnunu})
with the replacement $s\leftrightarrow d$.

\subsection*{Class VI: Lepton Number Violating Operators}

The operators that violate lepton number (but not baryon number) can be divided in two groups: operators with one charged lepton and a neutrino (Class VIa) 
and operators with two neutrinos (Class VIb and VIc). 
We use the notation $\psi^c = C \bar \psi^{\, \sss T}$ and $\bar \psi^{\, c} \equiv \psi^{\sss T} C$, where $C$ denotes the charge-conjugation matrix.

\medskip

\noindent $\blacktriangleright$ Class VIa : The following operators violate lepton number by $-2$ units.
\begin{align}
\O_{1}^{ub\ell\ell'^c} &= \left( \bar{u} \,P_R\, \gamma^\mu \, b \right)  \left( \bar{\ell} \, \gamma_\mu \,\nu_{\ell'}^c \right)\,, &
\O_{5}^{ub\ell\ell'^c} &= \left( \bar{u}\, P_R\, b \right)  \left( \bar{\ell}  \,\nu_{\ell'}^c \right)\,, &
\O_{7}^{ub\ell\ell'^c} &= \left( \bar{u} \, P_R\, \sigma^{\mu\nu}\,b \right)  \left( \bar{\ell} \, \sigma_{\mu\nu} \,\nu_{\ell'}^c \right), \notag \\[1mm]
\O_{1'}^{ub\ell\ell'^c} &= \left( \bar{u} \, P_L\, \gamma^\mu \, b \right)  \left( \bar{\ell} \, \gamma_\mu \,\nu_{\ell'}^c \right)\,, &
\O_{5'}^{ub\ell\ell'^c} &= \left( \bar{u}\, P_L \,b \right)  \left( \bar{\ell} \, \nu_{\ell'}^c \right)\,.
\end{align}
The corresponding $|\Delta C| =1$ operators $\O_i^{cb\ell\ell'^c}$ are obtained from $\O_i^{ub\ell\ell'^c}$ by performing the substitution $u\to c$.

\medskip

\noindent $\blacktriangleright$ Class VIb : The following operators violate lepton number by $-2$ units.
\begin{align}
  \O_{5}^{sb\ell\ell'^c} &= \left( \bar{s} \, P_R \, b \right) \left( \bar{\nu}_\ell \nu_{\ell'}^c \right), &
  \O_{7}^{sb\ell\ell'^c} &= \left( \bar{s} \,P_R\sigma^{\mu\nu}  \, b \right) \left( \bar{\nu}_\ell \sigma_{\mu\nu} \nu_{\ell'}^c \right),&
  \O_{5'}^{sb\ell\ell'^c} &= \left( \bar{s} \, P_L \, b \right) \left( \bar{\nu}_\ell \nu_{\ell'}^c \right).
  \label{eqn:sbvvLviolminus2}
\end{align}

\medskip

\noindent $\blacktriangleright$ Class VIc : The following operators violate lepton number by $+2$ units.
\begin{align}
  \O_{ 5}^{sb\ell^c\ell'}  &= \left( \bar{s} \, P_R \, b \right) \left( \bar{\nu}_\ell^{\,c} \nu_{\ell'} \right), &
  \O_{7'}^{sb\ell^c\ell'} &= \left( \bar{s} \,P_L\sigma^{\mu\nu}  \, b \right) \left( \bar{\nu}_\ell^{\,c} \sigma_{\mu\nu} \nu_{\ell'} \right),&
  \O_{5'}^{sb\ell^c\ell'} &= \left( \bar{s} \, P_L \, b \right) \left( \bar{\nu}_\ell^{\,c} \nu_{\ell'} \right).
  \label{eqn:sbvvLviolplus2}
\end{align}
In class VIb and VIc if we swap the generation indices $\ell \leftrightarrow \ell'$ the operators do not change since $(\bar{\nu}^{\,c}_i \, \nu_{j}) = (\bar{\nu}^{\,c}_{j} \, \nu_{i})$ and $(\bar{\nu}^{\,c}_i \, \sigma^{\mu\nu} \nu_{j}) = -(\bar{\nu}^{\,c}_{j} \, \sigma^{\mu\nu} \nu_{i})$. Therefore, for each possible $\ell\ell'$ pair only one must be considered.
The corresponding $|\Delta S| = 0 (|\Delta I| = 1/2)$ operators are obtained from Eqs.~\eqref{eqn:sbvvLviolminus2} and \eqref{eqn:sbvvLviolplus2} with the replacement $s \leftrightarrow d$.

\subsection*{Class VII : Baryon Number Violating  operators}

Baryon-number violating operators relevant for $B$ physics can be divided in two groups: operators that conserve $B-L$ (Classes VIIa - VIId) and operators that violate
$B-L$ (Classes VIIe - VIIi).\footnote{With $B$ here we mean \emph{baryon number}, not to be confused with \emph{beauty}, as in the rest of the paper.}
All operators violate also $L$: they contain either a charged lepton (Classes VIIa, VIIb, VIIe, VIIf and VIIg) or
 a neutrino (Classes VIIc, VIId, VIIh and VIIi).

\noindent $\blacktriangleright$ Class VIIa : 
\begin{align}
\O_1^{\ell bcu} &= \varepsilon_{\alpha\beta\lambda}\,
(\bar \ell^c\, P_R\, \gamma_\mu \,b_\alpha)\;   (\bar c^c_\beta \, \gamma^\mu\, u_\lambda)\,,&
\O_{1'}^{\ell bcu} &=\varepsilon_{\alpha\beta\lambda}\,
(\bar \ell^c \, P_L\, \gamma_\mu\, b_\alpha)\;   (\bar c^c_\beta\, \gamma^\mu\, u_\lambda)\,, \notag\\[1mm]
\O_3^{\ell bcu} &= \varepsilon_{\alpha\beta\lambda}\,
(\bar \ell^c\, P_R\, \gamma_{\mu\nu\rho} \,b_\alpha)\;   (\bar c^c_\beta \, \gamma^{\mu\nu\rho}\, u_\lambda)\,,&
\O_{3'}^{\ell bcu} &=\varepsilon_{\alpha\beta\lambda}\,
(\bar \ell^c \, P_L\, \gamma_{\mu\nu\rho}\, b_\alpha)\;   (\bar c^c_\beta\, \gamma^{\mu\nu\rho}\, u_\lambda)\,, \notag\\[1mm]
\O_5^{\ell bcu} &= \varepsilon_{\alpha\beta\lambda}\,
(\bar \ell^c\, P_R\, b_\alpha)\;   (\bar c^c_\beta \,  u_\lambda)\,,&
\O_{5'}^{\ell bcu} &=\varepsilon_{\alpha\beta\lambda}\,
(\bar \ell^c \, P_L\,  b_\alpha)\;   (\bar c^c_\beta\,  u_\lambda)\,, \\[1mm]
\O_7^{\ell bcu} &= \varepsilon_{\alpha\beta\lambda}\,
(\bar \ell^c\, P_R\, \sigma_{\mu\nu} \,b_\alpha)\;   (\bar c^c_\beta \, \sigma^{\mu\nu}\, u_\lambda)\,,&
\O_{7'}^{\ell bcu} &=\varepsilon_{\alpha\beta\lambda}\,
(\bar \ell^c \, P_L\, \sigma_{\mu\nu}\, b_\alpha)\;   (\bar c^c_\beta\, \sigma^{\mu\nu}\, u_\lambda)\,, \notag\\[1mm]
\O_9^{\ell bcu} &= \varepsilon_{\alpha\beta\lambda}\,
(\bar \ell^c\, P_R\, \gamma_{\mu\nu\rho\sigma} \,b_\alpha)\;   (\bar c^c_\beta \, \gamma^{\mu\nu\rho\sigma}\, u_\lambda)\,,&
\O_{9'}^{\ell bcu} &=\varepsilon_{\alpha\beta\lambda}\,
(\bar \ell^c \, P_L\, \gamma_{\mu\nu\rho\sigma}\, b_\alpha)\;   (\bar c^c_\beta\, \gamma^{\mu\nu\rho\sigma}\, u_\lambda)\,. \notag
\end{align}
Operators of the type
$\O_i^{\ell buc} = \varepsilon_{\alpha\beta\lambda}\,
(\bar \ell^c\, \Gamma_1^i\, b_\alpha)  (\bar u^c_\beta \,\Gamma_2^i \,  c_\lambda)$ are
all related to $\O_i^{\ell bcu}$ by transposition of the second current, and are not independent.

\medskip

\noindent $\blacktriangleright$ Class VIIb :
The cases $\O_i^{\ell buu}$ and $\O_i^{\ell bcc}$ are constrained by transpositions of the second current.
A set of independent operators is chosen as:
\begin{align}
\O_1^{\ell buu} &= \varepsilon_{\alpha\beta\lambda}\,
(\bar \ell^c\, P_R\, \gamma_\mu \,b_\alpha)\;   (\bar u^c_\beta \, \gamma^\mu\, u_\lambda)\,,&
\O_{1'}^{\ell buu} &=\varepsilon_{\alpha\beta\lambda}\,
(\bar \ell^c \, P_L\, \gamma_\mu\, b_\alpha)\;   (\bar u^c_\beta\, \gamma^\mu\, u_\lambda)\,, \notag\\[1mm]
\O_7^{\ell buu} &= \varepsilon_{\alpha\beta\lambda}\,
(\bar \ell^c\, P_R\, \sigma_{\mu\nu} \,b_\alpha)\;   (\bar u^c_\beta \, \sigma^{\mu\nu}\, u_\lambda)\,,&
\O_{7'}^{\ell buu} &=\varepsilon_{\alpha\beta\lambda}\,
(\bar \ell^c \, P_L\, \sigma_{\mu\nu}\, b_\alpha)\;   (\bar u^c_\beta\, \sigma^{\mu\nu}\, u_\lambda)\,.
\label{lbuu}
\end{align}
The corresponding operators $\O_i^{\ell bcc}$ are obtained from $\O_i^{\ell buu}$
by the substitution $u\to c$.

\medskip

\noindent $\blacktriangleright$ Class VIIc :
\begin{align}
  \O_{1'}^{\ell bus} &= \varepsilon_{\alpha\beta\lambda}\,
(\bar \nu_\ell^c\, P_L\, \gamma_\mu \,b_\alpha)\;   (\bar u^c_\beta \, \gamma^\mu\, s_\lambda)\,,&
\O_{3'}^{\ell bus} &= \varepsilon_{\alpha\beta\lambda}\,
(\bar \nu_\ell^c\, P_L\, \gamma_{\mu\nu\rho} \,b_\alpha)\;   (\bar u^c_\beta \, \gamma^{\mu\nu\rho}\, s_\lambda)\,, \notag\\[1mm]
\O_{5'}^{\ell bus} &= \varepsilon_{\alpha\beta\lambda}\,
(\bar \nu_\ell^c\, P_L\, b_\alpha)\;   (\bar u^c_\beta \,  s_\lambda)\,,&
\O_{7'}^{\ell bus} &= \varepsilon_{\alpha\beta\lambda}\,
(\bar \nu_\ell^c\, P_L\, \sigma_{\mu\nu} \,b_\alpha)\;   (\bar u^c_\beta \, \sigma^{\mu\nu}\, s_\lambda)\,, \notag\\[1mm]
\O_{9'}^{\ell bus} &= \varepsilon_{\alpha\beta\lambda}\,
(\bar \nu_\ell^c\, P_L\, \gamma_{\mu\nu\rho\sigma} \,b_\alpha)\;   (\bar u^c_\beta \, \gamma^{\mu\nu\rho\sigma}\, s_\lambda)\,.&
\label{lcbcs}
\end{align}
The corresponding operators $\O_{i'}^{\ell bud}$, $\O_{i'}^{\ell bcs}$ and $\O_{i'}^{\ell bcd}$
are obtained from $\O_{i'}^{\ell bus}$ by the substitution of the quark flavours $s\to d$, $u\to c$ and $(s,u)\to(d,c)$
respectively.

\medskip

\noindent $\blacktriangleright$ Class VIId :
The cases $\O_i^{\ell bub}$ and $\O_i^{\ell bcb}$ are constrained by Fierz identities;
here we choose the following minimal basis of independent operators:
\begin{align}
  \O_{1'}^{\ell bub} &= \varepsilon_{\alpha\beta\lambda}\,
(\bar \nu_\ell^c\, P_L\, \gamma_\mu \,b_\alpha)\;   (\bar u^c_\beta \, \gamma^\mu\, b_\lambda)\,,&
\O_{5'}^{\ell bub} &= \varepsilon_{\alpha\beta\lambda}\,
(\bar \nu_\ell^c\, P_L\, b_\alpha)\;   (\bar u^c_\beta \,  b_\lambda)\,.
\label{lcbcb}
\end{align}
In addition, $\O_i^{\ell bcb} = \O_i^{\ell bub} |_{u\to c}$.

\noindent The following classes correspond to $B-L$ violating operators.

\medskip

\noindent $\blacktriangleright$ Class VIIe :
The operators
\begin{align}
\O_1^{\ell bsd} &= \varepsilon_{\alpha\beta\lambda}\,
(\bar \ell\, P_R\, \gamma_\mu \,b_\alpha)\;   (\bar s^c_\beta \, \gamma^\mu\, d_\lambda)\,,&
\O_{1'}^{\ell bsd} &=\varepsilon_{\alpha\beta\lambda}\,
(\bar \ell \, P_L\, \gamma_\mu\, b_\alpha)\;   (\bar s^c_\beta\, \gamma^\mu\, d_\lambda)\,, \notag\\[1mm]
\O_3^{\ell bsd} &= \varepsilon_{\alpha\beta\lambda}\,
(\bar \ell\, P_R\, \gamma_{\mu\nu\rho} \,b_\alpha)\;   (\bar s^c_\beta \, \gamma^{\mu\nu\rho}\, d_\lambda)\,,&
\O_{3'}^{\ell bsd} &=\varepsilon_{\alpha\beta\lambda}\,
(\bar \ell \, P_L\, \gamma_{\mu\nu\rho}\, b_\alpha)\;   (\bar s^c_\beta\, \gamma^{\mu\nu\rho}\, d_\lambda)\,, \notag\\[1mm]
\O_5^{\ell bsd} &= \varepsilon_{\alpha\beta\lambda}\,
(\bar \ell\, P_R\, b_\alpha)\;   (\bar s^c_\beta \,  d_\lambda)\,,&
\O_{5'}^{\ell bsd} &=\varepsilon_{\alpha\beta\lambda}\,
(\bar \ell \, P_L\,  b_\alpha)\;   (\bar s^c_\beta\,  d_\lambda)\,, \\[1mm]
\O_7^{\ell bsd} &= \varepsilon_{\alpha\beta\lambda}\,
(\bar \ell\, P_R\, \sigma_{\mu\nu} \,b_\alpha)\;   (\bar s^c_\beta \, \sigma^{\mu\nu}\, d_\lambda)\,,&
\O_{7'}^{\ell bsd} &=\varepsilon_{\alpha\beta\lambda}\,
(\bar \ell \, P_L\, \sigma_{\mu\nu}\, b_\alpha)\;   (\bar s^c_\beta\, \sigma^{\mu\nu}\, d_\lambda)\,, \notag\\[1mm]
\O_9^{\ell bsd} &= \varepsilon_{\alpha\beta\lambda}\,
(\bar \ell\, P_R\, \gamma_{\mu\nu\rho\sigma} \,b_\alpha)\;   (\bar s^c_\beta \, \gamma^{\mu\nu\rho\sigma}\, d_\lambda)\,,&
\O_{9'}^{\ell bsd} &=\varepsilon_{\alpha\beta\lambda}\,
(\bar \ell \, P_L\, \gamma_{\mu\nu\rho\sigma}\, b_\alpha)\;   (\bar s^c_\beta\, \gamma^{\mu\nu\rho\sigma}\, d_\lambda)\,, \notag
\end{align}
mediate transitions of the type $\Lambda_b^0 \to K^+ e^-$.
The corresponding operators $\O_i^{\ell bds}$ are related to $\O_i^{\ell bsd}$
by transpositions of the second current and are not independent.

\medskip

\noindent $\blacktriangleright$ Class VIIf :
The cases $\O_i^{\ell bdd}$ and $\O_i^{\ell bss}$
(mediating transitions such as $\Lambda_b^0 \to \pi^+ e^-$ and $\Xi_b^0 \to K^+ e^-$ respectively)
are constrained by transpositions of the second current.
A set of independent operators is:
\begin{align}
\O_1^{\ell bss} &= \varepsilon_{\alpha\beta\lambda}\,
(\bar \ell\, P_R\, \gamma_\mu \,b_\alpha)\;   (\bar s^c_\beta \, \gamma^\mu\, s_\lambda)\,,&
\O_{1'}^{\ell bss} &=\varepsilon_{\alpha\beta\lambda}\,
(\bar \ell \, P_L\, \gamma_\mu\, b_\alpha)\;   (\bar s^c_\beta\, \gamma^\mu\, s_\lambda)\,, \notag\\[1mm]
\O_7^{\ell bss} &= \varepsilon_{\alpha\beta\lambda}\,
(\bar \ell\, P_R\, \sigma_{\mu\nu} \,b_\alpha)\;   (\bar s^c_\beta \, \sigma^{\mu\nu}\, s_\lambda)\,,&
\O_{7'}^{\ell bss} &=\varepsilon_{\alpha\beta\lambda}\,
(\bar \ell \, P_L\, \sigma_{\mu\nu}\, b_\alpha)\;   (\bar s^c_\beta\, \sigma^{\mu\nu}\, s_\lambda)\,,
\label{lbss}
\end{align}
with the corresponding operators $\O_i^{\ell bdd}$ obtained from $\O_i^{\ell bss}$
by the substitution $s\to d$.

\medskip

\noindent $\blacktriangleright$ Class VIIg :
The set $\O_i^{\ell bsb}$ (mediating, \emph{e.g.} $\Xi_b^0 \to B^+ e^-$) is constrained by Fierz identities,
resulting in only four independent operators. We choose:
\begin{align}
\O_1^{\ell bsb} &= \varepsilon_{\alpha\beta\lambda}\,
(\bar \ell\, P_R\, \gamma_\mu \,b_\alpha)\;   (\bar s^c_\beta \, \gamma^\mu\, b_\lambda)\,,&
\O_{1'}^{\ell bsb} &=\varepsilon_{\alpha\beta\lambda}\,
(\bar \ell \, P_L\, \gamma_\mu\, b_\alpha)\;   (\bar s^c_\beta\, \gamma^\mu\, b_\lambda)\,, \notag\\[1mm]
\O_7^{\ell bsb} &= \varepsilon_{\alpha\beta\lambda}\,
(\bar \ell\, P_R\, \sigma_{\mu\nu} \,b_\alpha)\;   (\bar s^c_\beta \, \sigma^{\mu\nu}\, b_\lambda)\,,&
\O_{7'}^{\ell bsb} &=\varepsilon_{\alpha\beta\lambda}\,
(\bar \ell \, P_L\, \sigma_{\mu\nu}\, b_\alpha)\;   (\bar s^c_\beta\, \sigma^{\mu\nu}\, b_\lambda)\,,
\label{lbsb}
\end{align}
and the operators $\O_i^{\ell bdb}$ are obtained by the substitution $s\to d$.
The operators $\O_i^{\ell bbs}$ and $\O_i^{\ell bbd}$ are related to the former by transposition of the second current,
and are not independent.

\medskip

\noindent $\blacktriangleright$ Class VIIh : The set of operators in class VIIh and VIIi correspond to the classes VIIc and VIId, respectively, where $\bar \nu_\ell^c$ is substituted with $\bar \nu_\ell$ and where the left-handed projector is interchanged with the right-handed one because of the opposite chirality of $\nu$ and $\nu^c$.
\begin{align}
\O_1^{\ell bus} &= \varepsilon_{\alpha\beta\lambda}\,
(\bar \nu_\ell\, P_R\, \gamma_\mu \,b_\alpha)\;   (\bar u^c_\beta \, \gamma^\mu\, s_\lambda)\,,&
\O_3^{\ell bus} &= \varepsilon_{\alpha\beta\lambda}\,
(\bar \nu_\ell\, P_R\, \gamma_{\mu\nu\rho} \,b_\alpha)\;   (\bar u^c_\beta \, \gamma^{\mu\nu\rho}\, s_\lambda)\,, \notag\\[1mm]
\O_5^{\ell bus} &= \varepsilon_{\alpha\beta\lambda}\,
(\bar \nu_\ell\, P_R\, b_\alpha)\;   (\bar u^c_\beta \,  s_\lambda)\,,&
\O_7^{\ell bus} &= \varepsilon_{\alpha\beta\lambda}\,
(\bar \nu_\ell\, P_R\, \sigma_{\mu\nu} \,b_\alpha)\;   (\bar u^c_\beta \, \sigma^{\mu\nu}\, s_\lambda)\,, \notag\\[1mm]
\O_9^{\ell bus} &= \varepsilon_{\alpha\beta\lambda}\,
(\bar \nu_\ell\, P_R\, \gamma_{\mu\nu\rho\sigma} \,b_\alpha)\;   (\bar u^c_\beta \, \gamma^{\mu\nu\rho\sigma}\, s_\lambda)\,.&
\label{lbcs}
\end{align}
The corresponding operators $\O_{i}^{\ell bud}$, $\O_{i}^{\ell bcs}$ and $\O_{i}^{\ell bcd}$ are obtained from $\O_{i}^{\ell bus}$ by the substitution of the quark flavours $s\to d$, $u\to c$ and $(s,u)\to(d,c)$ respectively.

\medskip

\noindent $\blacktriangleright$ Class VIIi :
\begin{align}
\O_1^{\ell bub} &= \varepsilon_{\alpha\beta\lambda}\,
(\bar \nu_\ell\, P_R\, \gamma_\mu \,b_\alpha)\;   (\bar u^c_\beta \, \gamma^\mu\, b_\lambda)\,,&
\O_5^{\ell bub} &= \varepsilon_{\alpha\beta\lambda}\,
(\bar \nu_\ell\, P_R\, b_\alpha)\;   (\bar u^c_\beta \,  b_\lambda)\,.
\label{lbcb}
\end{align}
In addition, $\O_{i}^{\ell bcb} = \O_{i}^{\ell bub} |_{u\to c}$.

\section{Renormalisation of the Effective Theory}
\label{renormalization}

The Wilson coefficients and dimension-six operators appearing in \Eq{LWET} are bare quantities and have to be renormalized.
The relationships between bare and renormalized quantities are given in terms of matrix-valued $\hat Z$ factors:
\eq{
\C_i^{\rm (0)} = Z_{ij}^c\,\C_j \ , \qquad \O_i^{\rm (0)} = Z_{ki}^\O\,\O_k\ .
}
The renormalization matrix $\hat Z_\O\equiv \hat Z^\O$ takes care of field renormalization, and possibly the renormalization of masses and
couplings that might appear in the normalization of the operators
(specifically in $\O_{7^{(\prime)}\gamma},\,\O_{8^{(\prime)}g}$). In our set-up $\hat Z_\O$ is always a diagonal matrix.
The renormalization matrix $\hat Z_c\equiv \hat Z^c$ takes care of the renormalization of the Wilson coefficients and
includes operator mixing. These renormalization factors depend on the renormalization scale and provide the renormalized
Wilson coefficients and operators with the corresponding renormalization scale dependence. In particular, since the bare
coefficients do not depend on the scale, one finds that (in matrix notation)
\eq{
\frac{d\,\vec \C }{d\log\mu} = -\hat Z_c^{-1} \frac{d \hat Z_c}{d\log\mu} \,\vec \C \ \equiv\ \hat \gamma^T\, \vec \C
}
which defines $\hat \gamma$, the anomalous dimension matrix.

The renormalization factors $\hat Z$ are calculated by subtracting the UV divergences of bare amplitudes perturbatively in a chosen renormalization scheme. 
In this paper we will regularize UV divergences by means of dimensional regularization in $D=4-2\epsilon$ dimensions, and subtract the divergences in the $\overline{\rm MS}$ scheme. However, the
one-loop anomalous dimensions will not depend on the renormalization scheme.\footnote{Barring the mass ratio issue mentioned below Eq.~\eqref{eqn:KJandM}.}
Scheme dependence only affects the finite one-loop terms, and all terms starting at two loops, which also depend on the choice of the evanescent operators.

Given the normalization of the operators in Section~\ref{set-up}, one loop corrections are always suppressed by
one power of $\alpha$, where $\alpha$ is either $\alpha_s$ or $\alpha_{\rm em}$ (the loop expansion coincides with the
coupling expansion). A generic renormalized amplitude can then be written as
\eqa{
\A &= &\bigg\{
\delta_{ij}
+ \frac{\alpha_s}{4\pi} \frac1{\hat \epsilon}
\Big[
\delta Z^{c,s}_{ij} + \delta Z^{\O,s}_{ij} + A^{s}_{ij}
\Big]
+ \frac{\alpha_{\rm em}}{4\pi} \frac1{\hat \epsilon}
\Big[
\delta Z^{c,{\rm em}}_{ij} + \delta Z^{\O,{\rm em}}_{ij} + A^{\rm em}_{ij}
\Big]
\bigg\}
\,\C_j \,\av{\O_i}^\text{tree} \nonumber \\[3mm]
&&+ \ \text{finite terms} \ + \O(\alpha_s^2, \alpha_{\rm em}^2, \alpha_s \alpha_{\rm em})\ ,
}
where the first two terms in each square bracket are the counterterm
contributions, the matrices $\hat A^s$, $\hat A^{\rm em}$  are the UV divergent
pieces of the bare one-loop amplitudes, and $\av{\O_i}^\text{tree}$
are the tree-level matrix elements of the operators. The scale dependence
is contained in the parameter
\eq{
\frac1{\hat \epsilon} = \frac1{\epsilon} - \gamma_E + \log(4\pi) - \log{\mu^2}\ .
}
The requirement that the one-loop divergences in the bare amplitudes are
cancelled by the counterterms leads to the equation:
\eq{
\hat \gamma^T =  \frac{\alpha_s}{4\pi} \Big[ -2 \hat A^{s} -2\,\delta \hat Z_{\O}^s \Big]
+\frac{\alpha_{\rm em}}{4\pi} \Big[ -2 \hat A^{\rm em} -2\,\delta \hat Z_\O^{\rm em} \Big]
+ \O(\alpha_s^2, \alpha_{\rm em}^2, \alpha_s \alpha_{\rm em})\ .
\label{ADMformula}
}
The renormalization factors $\delta \hat Z_{\O}$ are given by
\eq{
Z^\O_{ij} = \delta_{ij} + \frac{\alpha_s}{4\pi} \frac1{\hat \epsilon} \delta Z^{\O,s}_{ij} 
+ \frac{\alpha_{\rm em}}{4\pi} \frac1{\hat \epsilon} \delta Z^{\O,{\rm em}}_{ij}
+ \O(\alpha_s^2, \alpha_{\rm em}^2, \alpha_s \alpha_{\rm em})\ ,    
}
with
\eq{
\delta Z^{\O,s}_{ij} = \left\{
\begin{array}{rcl}
-2\,C_F\, \delta_{ij} &\quad& \text{for 4-quark operators} \\
-C_F\, \delta_{ij} && \text{for 2-quark-2-lepton operators} \\
\frac{1}{3}(11C_A-12C_F-4f\,T_F) && \text{for electromagnetic operators}\\
\frac{4}{3} \left(2 C_A-3 C_F-f \,T_F\right) && \text{for chromomagnetic operators}
\end{array}
\right.
}
with $C_A=3,\,C_F=4/3,\,T_F=1/2,\,f=5$, and
\eq{
\delta Z^{\O,{\rm em}}_{ij} = \left\{
\begin{array}{rcl}
-\frac12 (Q_1^2 + Q_2^2 + Q_3^2 + Q_4^2)\, \delta_{ij} &\quad& \text{for 4-fermion operators } (\bar\psi_1\Gamma \psi_2)(\bar\psi_3\Gamma' \psi_4) \\
-\frac{1}{2}(7Q_b^2+Q_s^2) && \text{for electromagnetic operators}\\
-\frac{1}{2}(7Q_b^2+Q_s^2) && \text{for chromomagnetic operators}
\end{array}
\right.
}

\begin{figure}
\centering
\subfloat[][]{\includegraphics[width=0.2\textwidth]{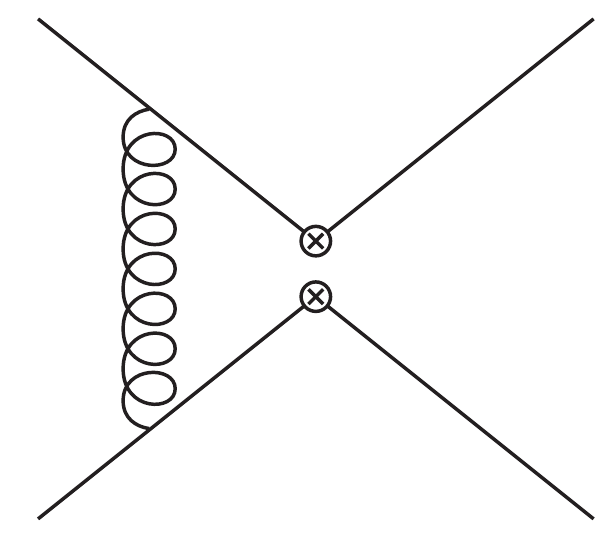}\label{fig:QCDvertexcorrection}}
\quad
\subfloat[][]{\includegraphics[width=0.2\textwidth]{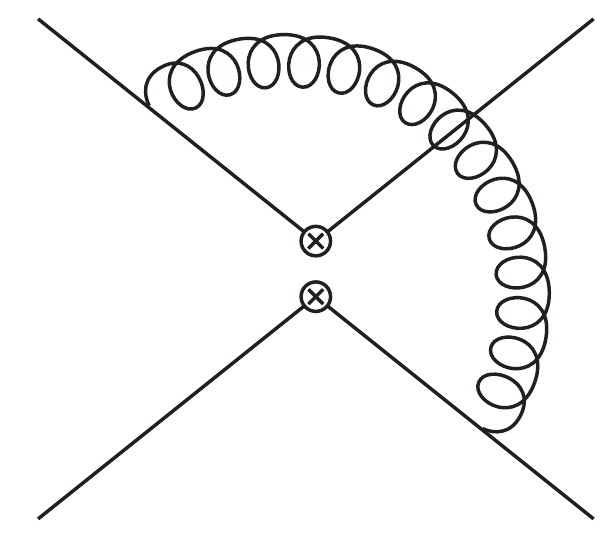}\label{fig:QCDvertexcorrection2}}
\quad
\subfloat[][]{\includegraphics[width=0.2\textwidth]{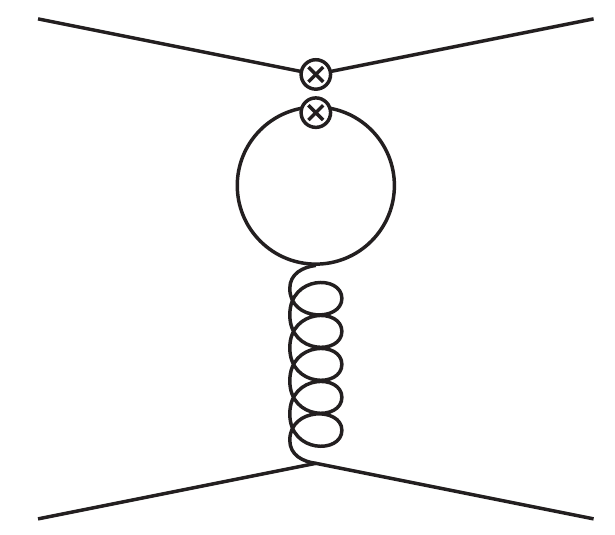}\label{fig:QCDclosedpenguin}}
\quad
\subfloat[][]{\includegraphics[width=0.2\textwidth]{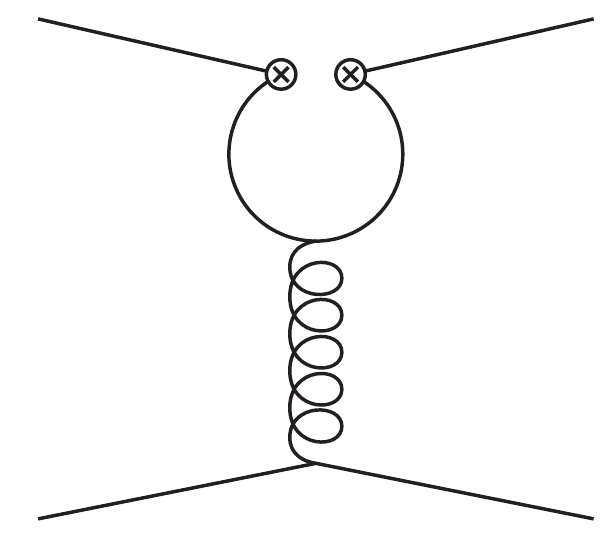}\label{fig:QCDopenpenguin}}
\\
\subfloat[][]{\includegraphics[width=0.2\textwidth]{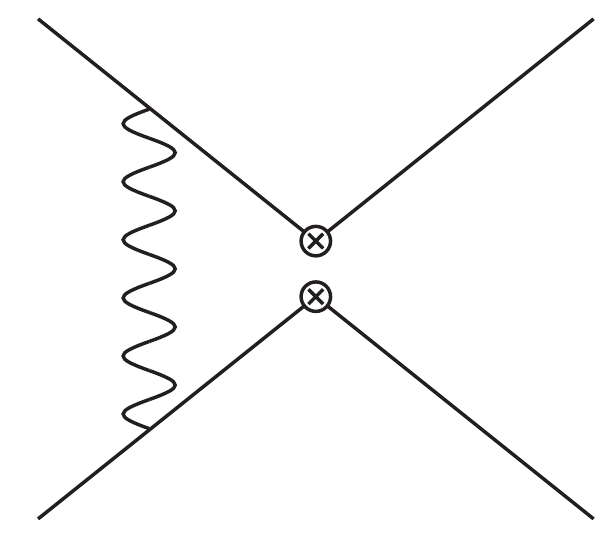}\label{fig:QEDvertexcorrection}}
\quad
\subfloat[][]{\includegraphics[width=0.2\textwidth]{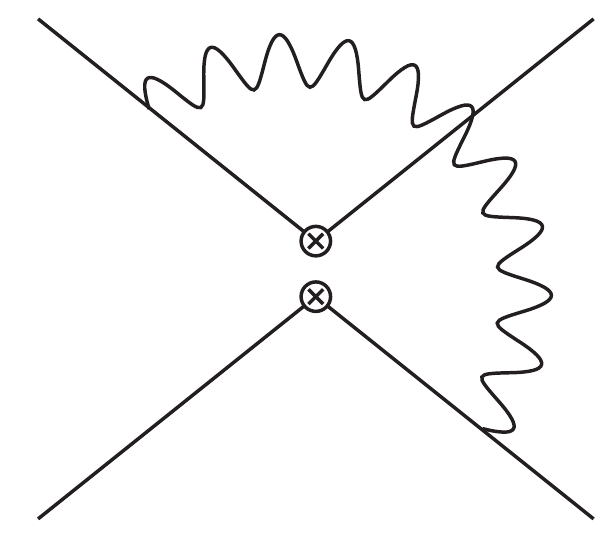}\label{fig:QEDvertexcorrection2}}
\quad
\subfloat[][]{\includegraphics[width=0.2\textwidth]{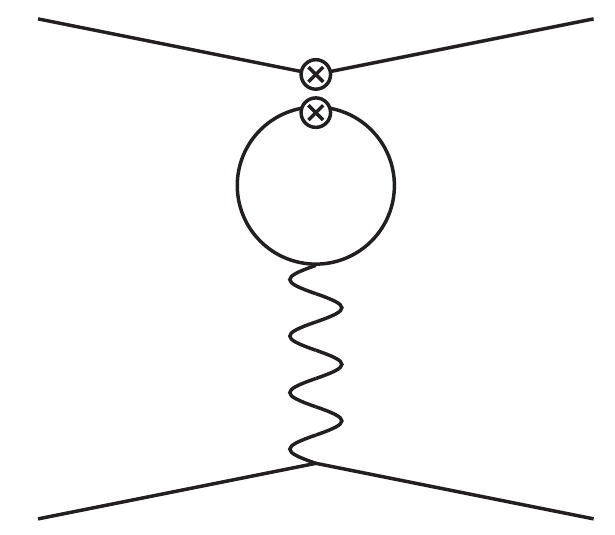}\label{fig:QEDclosedpenguin}}
\quad
\subfloat[][]{\includegraphics[width=0.2\textwidth]{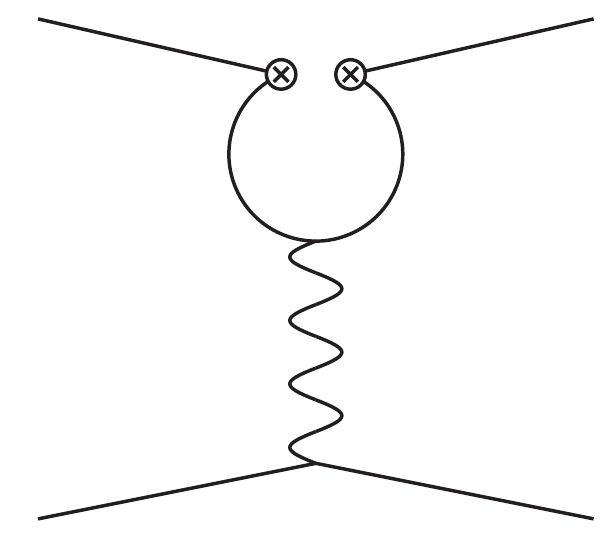}\label{fig:QEDopenpenguin}}
\caption{Representative set of one-loop penguin and vertex diagrams needed for the evaluation of the anomalous dimension
matrix at order $\alpha_s$ and $\alpha_{\rm em}$.}
\label{diagrams}
\end{figure}

The one-loop divergences in the bare amplitudes (the matrices $\hat A^{s}$ and $\hat A^{\rm em}$) are obtained by calculating
all one-loop QCD and QED corrections to the relevant amplitudes, expressing them in terms of tree level matrix
elements of the operators in the basis, and keeping only the $1/\epsilon$ poles. This requires the evaluation of
elementary one-loop penguin and vertex diagrams with one insertion of a dimension-six operator. A representative set
of the diagrams that have to be calculated is shown in Fig.~\ref{diagrams}.

\section{Complete Anomalous Dimensions Matrix at One Loop}
\label{ADM}

The complete one-loop ADM is obtained from \Eq{ADMformula} inserting the results for the $Z$ factors and one-loop
divergences outlined in Section~\ref{renormalization}. We have calculated all the entries of the ADM, and
compared our results for the entries that were already known, finding perfect agreement there. A summary
of pieces that were known and how to compare them to our results (in our new basis) is given in App.~\ref{changebasis}.

The full anomalous dimension matrix for the full set of operators listed in Table~\ref{TableList} has the following
block-diagonal form:
\eq{
\hat \gamma = {\rm Diagonal} \Big\{ \hat \gamma_{\rm \sss I},
\hat  \gamma_{\rm \sss II},
\hat  \gamma_{\rm \sss III},
\hat  \gamma_{\rm \sss IV},
\hat \gamma_{\rm \sss V},
\hat  \gamma_{\rm \sss Vb},
 \hat 0_{2\times 2},
\hat  \gamma_{\rm \sss VIa},
\hat  \gamma_{\rm \sss VIb},
\hat  \gamma_{\rm \sss VIc},
\hat  \gamma_{\rm \sss VIIa},\dots,
\hat  \gamma_{\rm \sss VIIi} \Big\}\ .
}
The different blocks $\hat \gamma_J$ have dimensions specified in Table~\ref{TableList}, and are given sequentially
in the remainder of this section.

\subsection*{Class I : $\mathbf{|\Delta B| = 2}$}

We combine all Class I operators into the following vector:
\eq{
\overrightarrow{\O_{\rm \sss I}} =
\{\O_1^{sbsb},\O_2^{sbsb},\O_3^{sbsb},\O_4^{sbsb},\O_5^{sbsb},\O_{1'}^{sbsb},\O_{2'}^{sbsb},\O_{3'}^{sbsb}\}.
}
The block $\hat \gamma_{\rm \sss I}$ in the order specified by $\overrightarrow{\O_{\rm \sss I}}$ is given by
\begin{equation}
\hat \gamma_{\rm \sss I}=\frac{\alpha_s}{4\pi} \begin{pmatrix}
4 & 0 & 0 & 0 & 0 & 0 & 0 & 0 \\
0 & -\frac{28}{3} & \frac{4}{3} & 0 & 0 & 0 & 0 & 0 \\
0 & \frac{16}{3} & \frac{32}{3} & 0 & 0 & 0 & 0 & 0 \\
0 & 0 & 0 & -16 & 0 & 0 & 0 & 0 \\
0 & 0 & 0 & -6 & 2 & 0 & 0 & 0 \\
0 & 0 & 0 & 0 & 0 & 4 & 0 & 0 \\
0 & 0 & 0 & 0 & 0 & 0 & -\frac{28}{3} & \frac{4}{3} \\
0 & 0 & 0 & 0 & 0 & 0 & \frac{16}{3} & \frac{32}{3} \\
\end{pmatrix}+
\frac{\alpha_{\text{em}}}{4\pi} \begin{pmatrix}
\frac{4}{3} & 0 & 0 & 0 & 0 & 0 & 0 & 0 \\
0 & -\frac{4}{9} & \frac{16}{9} & 0 & 0 & 0 & 0 & 0 \\
0 & \frac{16}{9} & -\frac{4}{9} & 0 & 0 & 0 & 0 & 0 \\
0 & 0 & 0 & -\frac{4}{3} & 0 & 0 & 0 & 0 \\
0 & 0 & 0 & 0 & -\frac{4}{3} & 0 & 0 & 0 \\
0 & 0 & 0 & 0 & 0 & \frac{4}{3} & 0 & 0 \\
0 & 0 & 0 & 0 & 0 & 0 & -\frac{4}{9} & \frac{16}{9} \\
0 & 0 & 0 & 0 & 0 & 0 & \frac{16}{9} & -\frac{4}{9} \\
\end{pmatrix}.
\label{eqn:ADMDeltaB2}
\end{equation}
The ADM corresponding to the set $\O_i^{dbdb}$ is identical.

\subsection*{Class II : $\mathbf{|\Delta B| = 1}$ semileptonic}

All Class II operators are combined into the vector:
\eq{
\overrightarrow{\O_{\rm \sss II}} =
\{\O_1^{ub\ell\ell'},O_5^{ub\ell\ell'},O_{1'}^{ub\ell\ell'},O_{5'}^{ub\ell\ell'},O_{7'}^{ub\ell\ell'}\} .
}
In this order, the block $\hat \gamma_{\rm \sss II}$ is given by:
\begin{equation}
\hat \gamma_{\rm \sss II}=\frac{\alpha_s}{4\pi} \begin{pmatrix}
0 & 0 & 0 & 0 & 0\\
0 & -8 & 0 & 0 & 0\\
0 & 0 & 0 & 0 & 0\\
0 & 0 & 0 & -8 & 0\\
0 & 0 & 0 & 0 & \frac83\\
\end{pmatrix} 
+\frac{\alpha_{\text{em}}}{4\pi} \begin{pmatrix}
-4 & 0 & 0 & 0 & 0\\
0  & \frac43 & 0 & 0 & 0\\
0  & 0 & -2 & 0 & 0\\
0  & 0 & 0 & \frac43 & \frac16\\
0  & 0 & 0 & 8 & -\frac{40}9\\
\end{pmatrix}.
\end{equation}
The ADM corresponding to the set $\O_i^{cb\ell\ell'}$ is identical.

\subsection*{Class III : $\mathbf{|\Delta B| = |\Delta C| =1}$  four-quark}

We group the unprimed Class III operators into the vector
\eq{
\overrightarrow{\O_{\rm \sss III}} =
\{\O_{1}^{sbuc},\O_{2}^{sbuc},\O_{3}^{sbuc},\cdots,\O_{9}^{sbuc},\O_{10}^{sbuc}  \} =
\{\O_{1-4}^{sbuc},\O_{6-10}^{sbuc}  \},
\label{OIII}
}
where in the second equality we have divided the set into two subsets.
With this notation, the block $\hat \gamma_{\rm \sss III}$ has itself a sub-block-diagonal form:
\begin{equation}
   \hat \gamma_{\rm \sss III} =
   \begin{pmatrix}
     \hat \Gamma^{1-4}_{\rm \sss III} & 0 \\
     0 & \hat \Gamma^{5-10}_{\rm \sss III} \\
   \end{pmatrix},
\label{gammaIII}
\end{equation}
with the following sub-blocks:
\begin{align}
 \hat \Gamma^{1-4}_{\rm \sss III} &=\frac{\alpha_s}{4\pi} 
  \begin{pmatrix}
     0 & -20 & 0 & 2 \\
     -\frac{40}{9} & -\frac{52}{3} & \frac{4}{9} & \frac{5}{6} \\
     0 & -128 & 0 & 20 \\
     -\frac{256}{9} & -\frac{160}{3} & \frac{40}{9} & -\frac{2}{3} \\
  \end{pmatrix}+\frac{\alpha_{\text{em}}}{4\pi} 
  \begin{pmatrix}
     \frac{40}{9} & 0 & -\frac{4}{9} & 0 \\
     0 & \frac{40}{9} & 0 & -\frac{4}{9} \\
     \frac{256}{9} & 0 & -\frac{40}{9} & 0 \\
     0 & \frac{256}{9} & 0 & -\frac{40}{9} \\
  \end{pmatrix}, \\\nonumber
\\
\hat \Gamma^{5-10}_{\rm \sss III} &=\frac{\alpha_s}{4\pi}
  \begin{pmatrix}
    -16 & 0 & 0 & -2 & 0 & 0 \\
     0 & 2 & -\frac{4}{9} & -\frac{5}{6} & 0 & 0 \\
     0 & 32 & \frac{16}{3} & -32 & 0 & -2 \\
     \frac{64}{9} & \frac{40}{3} & -\frac{64}{9} & -26 & -\frac{4}{9} & -\frac{5}{6} \\
     0 & -512 & -\frac{1024}{3} & 384 & -16 & 32 \\
     -\frac{1024}{9} & -\frac{640}{3} & \frac{256}{3} & \frac{1184}{3} & \frac{64}{9} & \frac{46}{3}  &
  \end{pmatrix}
+ \frac{\alpha_{\text{em}}}{4\pi}
 \begin{pmatrix}
   -\frac{10}{3} & 0 & \frac{4}{9} & 0 & 0 & 0 \\
    0 & -\frac{10}{3} & 0 & \frac{4}{9} & 0 & 0 \\
    -\frac{64}{9} & 0 & \frac{74}{9} & 0 & \frac{4}{9} & 0 \\
    0 & -\frac{64}{9} & 0 & \frac{74}{9} & 0 & \frac{4}{9} \\
    \frac{1024}{9} & 0 & -\frac{1408}{9} & 0 & -\frac{94}{9} & 0 \\
    0 & \frac{1024}{9} & 0 & -\frac{1408}{9} & 0 & -\frac{94}{9} \\ 
 \end{pmatrix}.
\end{align}
The anomalous dimensions for the set of primed operators $\O_{i'}^{sbuc}$
are identical, as well as the ones for the sets $\O_{i}^{dbuc}$,
$\O_{i}^{sbcu}$ and $\O_{i}^{dbcu}$ and their primed counterparts.

\subsection*{Class IV : $\mathbf{|\Delta B| = 1,\ |\Delta S| =2}$  four-quark}

The operators in Class IV are ordered and grouped into the following vector:
\eq{
\overrightarrow{\O_{\rm \sss IV}} =
\{\O_1^{sbsd},\O_3^{sbsd},\O_5^{sbsd},\O_7^{sbsd},\O_9^{sbsd}\},
}
with respect to which the block $\hat \gamma_{\rm \sss IV}$ is given by:
\begin{equation}
\hat \gamma_{\rm \sss IV}=\frac{\alpha_s}{4\pi} \begin{pmatrix}
\frac{4}{3} & \frac{1}{6} & 16 & -4 & -\frac{1}{4} \\
 -\frac{32}{3} & \frac{14}{3} & 64 & -16 & -1 \\
 0 & 0 & -18 & \frac{11}{6} & \frac{1}{8} \\
 0 & 0 & -\frac{40}{3} & \frac{74}{3} & \frac{5}{6} \\
 0 & 0 & \frac{256}{3} & -\frac{1600}{3} & -\frac{64}{3} \\
\end{pmatrix} 
+\frac{\alpha_{\text{em}}}{4\pi} \begin{pmatrix}
-\frac{20}{9} & \frac{2}{9} & 0 & 0 & 0 \\
 -\frac{128}{9} & \frac{20}{9} & 0 & 0 & 0 \\
 0 & 0 & -\frac{4}{3} & -\frac{2}{9} & 0 \\
 0 & 0 & \frac{32}{9} & -\frac{28}{9} & -\frac{2}{9} \\
 0 & 0 & -\frac{512}{9} & \frac{128}{9} & \frac{20}{9} \\
\end{pmatrix}.
\end{equation}
The anomalous dimensions for the set of primed operators $\O_{i'}^{sbsd}$
are identical, as well as the ones for the set $\O_{i}^{dbds}$,
and its primed counterpart.

\subsection*{Class V : $\mathbf{|\Delta B| = 1,\ |\Delta C| = 0}$}

The block $\hat \gamma_{\rm \sss V}$ is the largest one, given by a $57\times 57$
matrix (plus an identical copy for the primed operators). This block can itself be
divided in sub-blocks, which is instructive since this already unfolds most of the features of the mixing pattern.
We order the complete basis of Class~V operators into the vector:
\eq{
\overrightarrow{\O_{\rm \sss V}} = \Big\{
\O^{sbuu}_{1\mhyphen4},
\O^{sbuu}_{5\mhyphen10},
\O^{sbdd}_{1\mhyphen4},
\O^{sbdd}_{5\mhyphen10},
\O^{sbcc}_{1\mhyphen4}, 
\O^{sbcc}_{5\mhyphen10},
\O^{sbss}_{1\mhyphen9}, 
\O^{sbbb}_{1\mhyphen9},
\O^s_{7\gamma,8g}, 
\O_{1-9}^{sbee},
O_{1-9}^{sb\mu\mu},
O_{1-9}^{sb\tau\tau} 
\Big\}
}
which defines also the different sub-blocks in the matrix. Then,
\eq{
\renewcommand{\arraystretch}{1.1}
\arraycolsep=7pt
\hat \gamma_{\rm \sss V} = \left(
\begin{array}{cc|cc|cc|c|c|c|c|c|c}
A^u &   & Z^d &   & Z^u &   & H^u & H^u &   &  N^u    &  N^u & N^u        \\
& B^u &   &   &   &   &   &   &   &   &   &             \\
\hline
Z^d &   & A^d &   & Z^d &   & H^d & H^d &   & N^d & N^d & N^d   \\
&   &   & B^d &   &   &   &   &   &   &   &          \\
\hline
Z^u &   & Z^d &   & A^u &   & H^u & H^u &   & N^u   & N^u  & N^u      \\
&   &   &   &   & B^u &   &   &  K &   &   &        \\
\hline
I^u &   & I^d &   & I^u &   & C & D &   & P & P  & P   \\
\hline
I^u &   & I^d &   & I^u &   & D & C & J  & P & P & P  \\
\hline
&   &   &   &   &   &   &   & E &   &   &       \\
\hline
L^u  &   & L^d  &   &  L^u &   & Q  & Q  &   & F &  G    &  G   \\
\hline
L^u  &   &  L^d &   &  L^u &   &  Q & Q  &   &  G    & F    & G  \\
\hline
L^u  &   & L^d  &   &  L^u &   & Q  & Q  & M  &  G    &  G    & F 
\end{array}
\right),
\label{eqn:ADMclassV}
}
where the empty entries represent zeroes. The different sub-blocks are
as follows:

\allowdisplaybreaks

\noindent The diagonal entries are given by:
\begin{align}
 \hat A^q&=\frac{\alpha_s}{4\pi} 
  \begin{pmatrix}
     0 & -20 & 0 & 2 \\
     -\frac{40}{9} & -16 & \frac{4}{9} & \frac{5}{6} \\
     0 & -128 & 0 & 20 \\
     -\frac{256}{9} & -40 & \frac{40}{9} & -\frac{2}{3} \\
  \end{pmatrix}+\frac{\alpha_{\text{em}}}{4\pi}Q_q\left(
\begin{array}{cccc}
 \frac{20}{3}+8Q_q & 0 & -\frac{2}{3} & 0 \\
 0 & \frac{20}{3} & 0 & -\frac{2 }{3} \\
 \frac{128 }{3}+80Q_q & 0 & -\frac{20 }{3} & 0 \\
 0 & \frac{128 }{3} & 0 & -\frac{20 }{3} \\
\end{array}
\right), \\\nonumber
\\
\hat B^u&=\frac{\alpha_s}{4\pi}
  \begin{pmatrix}
    -16 & 0 & 0 & -2 & 0 & 0 \\
     0 & 2 & -\frac{4}{9} & -\frac{5}{6} & 0 & 0 \\
     0 & 32 & \frac{16}{3} & -32 & 0 & -2 \\
     \frac{64}{9} & \frac{40}{3} & -\frac{64}{9} & -26 & -\frac{4}{9} & -\frac{5}{6} \\
     0 & -512 & -\frac{1024}{3} & 384 & -16 & 32 \\
     -\frac{1024}{9} & -\frac{640}{3} & \frac{256}{3} & \frac{1184}{3} & \frac{64}{9} & \frac{46}{3}  &
 \end{pmatrix}+\frac{\aem}{4\pi}
 \begin{pmatrix}
   -\frac{10}{3} & 0 & \frac{4}{9} & 0 & 0 & 0 \\
    0 & -\frac{10}{3} & 0 & \frac{4}{9} & 0 & 0 \\
    -\frac{64}{9} & 0 & \frac{74}{9} & 0 & \frac{4}{9} & 0 \\
    0 & -\frac{64}{9} & 0 & \frac{74}{9} & 0 & \frac{4}{9} \\
    \frac{1024}{9} & 0 & -\frac{1408}{9} & 0 & -\frac{94}{9} & 0 \\
    0 & \frac{1024}{9} & 0 & -\frac{1408}{9} & 0 & -\frac{94}{9}  &
 \end{pmatrix},\\
\hat B^d&=\frac{\alpha_s}{4\pi}
  \begin{pmatrix}
    -16 & 0 & 0 & -2 & 0 & 0 \\
     0 & 2 & -\frac{4}{9} & -\frac{5}{6} & 0 & 0 \\
     0 & 32 & \frac{16}{3} & -32 & 0 & -2 \\
     \frac{64}{9} & \frac{40}{3} & -\frac{64}{9} & -26 & -\frac{4}{9} & -\frac{5}{6} \\
     0 & -512 & -\frac{1024}{3} & 384 & -16 & 32 \\
     -\frac{1024}{9} & -\frac{640}{3} & \frac{256}{3} & \frac{1184}{3} & \frac{64}{9} & \frac{46}{3}  &
 \end{pmatrix}+\frac{\aem}{4\pi}
 \begin{pmatrix}
   -\frac{4}{3} & 0 & -\frac{2}{9} & 0 & 0 & 0 \\
    0 & -\frac{4}{3} & 0 & -\frac{2}{9} & 0 & 0 \\
    \frac{32}{9} & 0 & -\frac{28}{9} & 0 & -\frac{2}{9} & 0 \\
    0 & \frac{32}{9} & 0 & -\frac{28}{9} & 0 & -\frac{2}{9} \\
    -\frac{512}{9} & 0 & \frac{128}{9} & 0 & \frac{20}{9} & 0 \\
    0 & -\frac{512}{9} & 0 & \frac{128}{9} & 0 & \frac{20}{9} \\ 
 \end{pmatrix},\\
  \hat C&=\frac{\alpha_s}{4\pi}
  \begin{pmatrix}
    \frac{8}{9} & \frac{2}{9} & \frac{128}{9} & -\frac{32}{9} & -\frac{2}{9} \\
   -\frac{160}{9} & \frac{50}{9} & \frac{320}{9} & -\frac{80}{9} & -\frac{5}{9} \\
    \frac{2}{9} & -\frac{1}{36} & -\frac{154}{9} & \frac{29}{18} & \frac{1}{9} \\
    0 & 0 & -\frac{40}{3} & \frac{74}{3} & \frac{5}{6} \\
    \frac{32}{9} & -\frac{4}{9} & \frac{896}{9} & -\frac{4832}{9} & -\frac{194}{9} \\ 
  \end{pmatrix}+\frac{\aem}{4\pi}
  \begin{pmatrix}
    -\frac{32}{27} & \frac{2}{9} & 0 & 0 & 0 \\
   -\frac{80}{27} & \frac{20}{9} & 0 & 0 & 0 \\
    -\frac{2}{27} & 0 & -\frac{4}{3} & -\frac{2}{9} & 0 \\
    0 & 0 & \frac{32}{9} & -\frac{28}{9} & -\frac{2}{9} \\
    -\frac{32}{27} & 0 & -\frac{512}{9} & \frac{128}{9} & \frac{20}{9} \\ 
  \end{pmatrix}, \\\nonumber
\\
 \hat E&=
  \frac{\alpha_s}{4\pi}
  \left(
\begin{array}{cc}
 -\frac{14}{3} & 0 \\
 -\frac{32}{9} & -6 \\
\end{array}
\right)
+\frac{\aem}{4\pi}
  \left(
\begin{array}{cc}
 \frac{16}{9} & -\frac{8}{3} \\
 0 & \frac{8}{9} \\
\end{array}
\right), \\\nonumber
\\
 \hat  F&=  \frac{\alpha_s}{4\pi}\begin{pmatrix}
      0 & 0 & 0 & 0 & 0\\
      0 & 0 & 0 & 0 & 0\\
      0 & 0 & -8 & 0 & 0\\
      0 & 0 & 0 & \frac{8}{3} & 0\\
      0 & 0 & 0 & -\frac{512}{3} & -8\\
  \end{pmatrix}+
\frac{\aem}{4\pi}\begin{pmatrix}
      -4 & \frac{2}{3} & 0 & 0 & 0\\
      -16 & \frac{20}{3} & 0 & 0 & 0\\
      0 & 0 & -\frac{20}{3} & -\frac{2}{3} & 0\\
      0 & 0 & \frac{32}{3} & -\frac{76}{9} & -\frac{2}{3}\\
      0 & 0 & - \frac{512}{3} & -\frac{128}{9} & 4\\
  \end{pmatrix}.
\end{align}

\noindent The mixing among the four-quark operators is given by the following matrices:
\begin{align}
\hat  D&=\frac{\alpha_s}{4\pi}
  \begin{pmatrix}
    -\frac{4}{9} & \frac{1}{18} & -\frac{16}{9} & \frac{4}{9} & \frac{1}{36} \\
    -\frac{64}{9} & \frac{8}{9} & -\frac{256}{9} & \frac{64}{9} & \frac{4}{9} \\
    \frac{2}{9} & -\frac{1}{36} & \frac{8}{9} & -\frac{2}{9} & -\frac{1}{72} \\
    0 & 0 & 0 & 0 & 0 \\
    \frac{32}{9} & -\frac{4}{9} & \frac{128}{9} & -\frac{32}{9} & -\frac{2}{9} \\ 
  \end{pmatrix} +
\frac{\aem}{4\pi}
\left(
\begin{array}{ccccc}
 \frac{28}{27} & 0 & 0 & 0 & 0 \\
 \frac{304}{27} & 0 & 0 & 0 & 0 \\
 -\frac{2}{27} & 0 & 0 & 0 & 0 \\
 0 & 0 & 0 & 0 & 0 \\
 -\frac{32}{27} & 0 & 0 & 0 & 0 \\
\end{array}
\right),
\\
\hat Z^q&=\frac{\alpha_s}{4\pi}
  \begin{pmatrix}
     0 & 0 & 0 & 0 \\
     0 & \frac{4}{3} & 0 & 0 \\
     0 & 0 & 0 & 0 \\
     0 & \frac{40}{3} & 0 & 0 \\
  \end{pmatrix}+\frac{\aem}{4\pi}Q_q\left(
\begin{array}{cccc}
 \frac{16}{3} & 0 & 0 & 0 \\
 0 & 0 & 0 & 0 \\
 \frac{160}{3} & 0 & 0 & 0 \\
 0 & 0 & 0 & 0 \\
\end{array}
\right) , \\
\hat I^q & =\frac{\alpha_s}{4\pi}
  \begin{pmatrix}
      0 & \frac{4}{3} & 0 & 0 \\
      0 & \frac{64}{3} & 0 & 0 \\
      0 & -\frac{2}{3} & 0 & 0 \\
      0 & 0 & 0 & 0 \\
      0 & -\frac{32}{3} & 0 & 0 \\
  \end{pmatrix}+\frac{\aem}{4\pi}Q_q\left(
\begin{array}{cccc}
 -\frac{28}{9} & 0 & 0 & 0 \\
 -\frac{304}{9} & 0 & 0 & 0 \\
 \frac{2}{9} & 0 & 0 & 0 \\
 0 & 0 & 0 & 0 \\
 \frac{32}{9} & 0 & 0 & 0 \\
\end{array}
\right),\\
\hat H^q&=\frac{\alpha_s}{4\pi}
  \begin{pmatrix} 
     0 & 0 & 0 & 0 & 0 \\
     -\frac{4}{9} & \frac{1}{18} & -\frac{16}{9} & \frac{4}{9} & \frac{1}{36} \\
     0 & 0 & 0 & 0 & 0 \\
     -\frac{40}{9} & \frac{5}{9} & -\frac{160}{9} & \frac{40}{9} & \frac{5}{18} \\
  \end{pmatrix}+\frac{\aem}{4\pi}Q_q\left(
\begin{array}{ccccc}
 -\frac{8}{3} & 0 & 0 & 0 & 0 \\
 0 & 0 & 0 & 0 & 0 \\
 -\frac{80}{3} & 0 & 0 & 0 & 0 \\
 0 & 0 & 0 & 0 & 0 \\
\end{array}
\right).
\end{align}

\noindent The matrices describing the mixing of four-fermion operators into electro- and chromomagnetic operators only contain an $\alpha_s$ part, due to the normalization of $\O^s_{7\gamma},\O^s_{8g}$. They are given by:
\eq{
\hat K=
\frac{\alpha_s}{4\pi}
\left(
\begin{array}{cc}
 0 & 0 \\
 0 & 0 \\
 -16 x_c & 0 \\
 0 & -4 x_c \\
 256 x_c & 0 \\
 0 & 64 x_c \\
\end{array}
\right)\ ,\quad
\hat J=  \frac{\alpha_s}{4\pi}
\left(
\begin{array}{cc}
 0 & 0 \\
 0 & 0 \\
 -\frac{1}{3} & 1 \\
 \frac{28}{3} & -4 \\
 -\frac{512}{3} & 128 \\
\end{array}
\right)\ ,\quad
\hat M=\frac{\alpha_s}{4\pi}
\left(
\begin{array}{cc}
 0 & 0 \\
 0 & 0 \\
 0 & 0 \\
 8\,x_{\tau} & 0 \\
 -128\,x_{\tau} & 0 \\
\end{array}
\right).
\label{eqn:KJandM}
}
The matrices $\hat K$ and $\hat M$ depend on the parameters $x_c \equiv m_c/m_b$ and $x_\tau 
\equiv m_\tau/m_b$, respectively, that make the ADM scale- and scheme-dependent.
However such dependence can be removed in principle by including in Eq.~\eqref{sbM} new dipole operators normalized with $m_c$ or $m_\tau$ instead of $m_b$.

\noindent The mixing among operators containing different leptonic flavours is given by:
\begin{equation}
\hat  G=  \frac{\aem}{4\pi}\begin{pmatrix}
      \frac{8}{3} & 0 & 0 & 0 & 0\\
      \frac{80}{3} & 0 & 0 & 0 & 0\\
      0 & 0 & 0 & 0 & 0\\
      0 & 0 & 0 & 0 & 0\\
      0 & 0 & 0 & 0 & 0\\
  \end{pmatrix}.
\end{equation}

\noindent The mixing of the semileptonic operators into four-quark operators is given by:
\begin{align}
\hat L^q&=\frac{\aem}{4\pi}
Q_q\left(
\begin{array}{cccc}
 -\frac{8 }{3} & 0 & 0 & 0 \\
 -\frac{80 }{3} & 0 & 0 & 0 \\
  0 & 0 & 0 & 0 \\
 0 & 0 & 0 & 0 \\
 0 & 0 & 0 & 0 \\
\end{array}
\right),&
\hat Q=\frac{\aem}{4\pi}
\left(
\begin{array}{ccccc}
 \frac{8}{9} & 0 & 0 & 0 & 0 \\
 \frac{80}{9} & 0 & 0 & 0 & 0 \\
 0 & 0 & 0 & 0 & 0 \\
 0 & 0 & 0 & 0 & 0 \\
 0 & 0 & 0 & 0 & 0 \\
\end{array}
\right)\ .
\end{align}

\noindent The mixing of four-quark operators into semileptonic operators is given by:
\begin{align}
\hat N^q&=\frac{\aem}{4\pi}
Q_q\left(
\begin{array}{ccccc}
 -8 & 0 & 0 & 0 & 0 \\
 0 & 0 & 0 & 0 & 0 \\
 -80 & 0 & 0 & 0 & 0 \\
 0 & 0 & 0 & 0 & 0 \\
\end{array}
\right),&
\hat P=\frac{\aem}{4\pi}
\left(
\begin{array}{ccccc}
 \frac{28}{9} & 0 & 0 & 0 & 0 \\
 \frac{304}{9} & 0 & 0 & 0 & 0 \\
 -\frac{2}{9} & 0 & 0 & 0 & 0 \\
 0 & 0 & 0 & 0 & 0 \\
 -\frac{32}{9} & 0 & 0 & 0 & 0 \\
\end{array}
\right).
\end{align}

\noindent Finally, for the lepton-flavour violating operators in {\bf Class Vb}, $\{\O_{1-9}^{bs\ell\ell'}\}$ with $\ell\neq \ell'$ we find:
\begin{equation}
   \hat \gamma_{\rm \sss Vb} =
   \frac{\alpha_s}{4\pi}\begin{pmatrix}
      0 & 0 & 0 & 0 & 0\\
      0 & 0 & 0 & 0 & 0\\
      0 & 0 & -8 & 0 & 0\\
      0 & 0 & 0 & \frac{8}{3} & 0\\
      0 & 0 & 0 & -\frac{512}{3} & -8\\
  \end{pmatrix}+\frac{\aem}{4\pi}\left(
\begin{array}{ccccc}
 -\frac{20}{3} & \frac{2}{3} & 0 & 0 & 0 \\
 -\frac{128}{3} & \frac{20}{3} & 0 & 0 & 0 \\
 0 & 0 & -\frac{20}{3} & -\frac{2}{3} & 0 \\
 0 & 0 & \frac{32}{3} & -\frac{76}{9} & -\frac{2}{3} \\
 0 & 0 & -\frac{512}{3} & -\frac{128}{9} & 4 \\
\end{array}
\right)\ .
\label{eqn:ADMDeltaB2llprime}
\end{equation} \\                                                  
All these matrices replicate exactly for the corresponding sets of primed operators,
as well as for the operators mediating $b\to d$ transitions. We also reiterate that
the ADM for the Class V$\nu$ operators $\O_{\nu 1,1'}^{sb\ell\ell'}$ vanishes.

\subsection*{Class VI : Lepton Number Violating}

We group the Class VI operators in the following way:
\begin{align}
\overrightarrow{\O_{\rm \sss VIa}} &=
 \{\O_1^{ub\ell\ell'^c},O_5^{ub\ell\ell'^c},O_{7}^{ub\ell\ell'^c},O_{1'}^{ub\ell\ell'^c},O_{5'}^{ub\ell\ell'^c}\}\, , & 
 \overrightarrow{\O_{\rm \sss VIb}} &= 
 \{\O_5^{sb\ell\ell'^c}, \O_7^{sb\ell\ell'^c}, \O_{5'}^{sb\ell\ell'^c} \}\, ,\notag \\[1mm]
 \overrightarrow{\O_{\rm \sss VIc}} &= 
 \{\O_{5'}^{sb\ell^c \ell'} , \O_{7'}^{sb\ell^c \ell'}, \O_{5}^{sb\ell^c\ell'} \}\, . 
\end{align}
The blocks $\hat \gamma_{\rm\sss VIa-c}$ are given by:
\begin{align}
\hat \gamma_{\rm \sss VIa} &= \frac{\alpha_s}{4 \pi}
\begin{pmatrix}
0 & 0 & 0 & 0 & 0 \\ 
0 & -8 & 0 & 0 & 0 \\ 
0 & 0 & \frac{8}{3} & 0 & 0 \\ 
0 & 0 & 0 & 0 & 0 \\ 
0 & 0 & 0 & 0 & -8
\end{pmatrix}
+
\frac{\alpha_{\rm\sss em}}{4\pi}
\begin{pmatrix}
-2 & 0 & 0 & 0 & 0 \\ 
0 & \frac{4}{3} & \frac{1}{6} & 0 & 0 \\ 
0 & 8 & -\frac{40}{9} & 0 & 0 \\ 
0 & 0 & 0 & -4 & 0 \\ 
0 & 0 & 0 & 0 & \frac{4}{3}
\end{pmatrix}\, ,\\[1mm]
\hat \gamma_{\rm \sss VIb} &=  \hat \gamma_{\rm \sss VIc} =
\frac{\alpha_s}{4 \pi}
\begin{pmatrix}
-8 & 0 & 0 \\ 
0 & \frac{8}{3} & 0 \\ 
0 & 0 & -8
\end{pmatrix} 
+
\frac{\alpha_{\rm\sss em}}{4\pi}
\begin{pmatrix}
-\frac{2}{3} & 0 & 0 \\ 
0 & \frac{2}{9} & 0 \\ 
0 & 0 & -\frac{2}{3}
\end{pmatrix}\, .
\end{align}

\subsection*{Class VII : Baryon Number Violating}

We define the following vectors for the Class VII operators:
\begin{align}
\overrightarrow{\O_{\rm \sss VIIa}} &= \{\O_1^{\ell bcu},\O_3^{\ell bcu},\O_5^{\ell bcu},\O_7^{\ell bcu},\O_9^{\ell bcu}\}\ ,
\quad&
\overrightarrow{\O_{\rm \sss VIIb}} &= \{\O_1^{\ell buu},\O_7^{\ell buu}  \}\ ,  \notag \\[1mm]
\overrightarrow{\O_{\rm \sss VIIc}} &= \{\O_{1'}^{\ell bus},\O_{3'}^{\ell bus},\O_{5'}^{\ell bus},\O_{7'}^{\ell bus},\O_{9'}^{\ell bus}\}\ ,
\quad&
\overrightarrow{\O_{\rm \sss VIId}} &= \{\O_{1'}^{\ell bub},\O_{5'}^{\ell bub}  \}\ ,  \\[1mm]
\overrightarrow{\O_{\rm \sss VIIe}} &= \{\O_1^{\ell bsd},\O_3^{\ell bsd},\O_5^{\ell bsd},\O_7^{\ell bsd},\O_9^{\ell bsd}\}\ , 
\quad&
\overrightarrow{\O_{\rm \sss VIIf}} &= \{\O_1^{\ell bss},\O_7^{\ell bss}  \}\ , \notag\\[1mm]
\overrightarrow{\O_{\rm \sss VIIg}} &= \{\O_1^{\ell bsb},\O_7^{\ell bsb}  \}\ , & 
\overrightarrow{\O_{\rm \sss VIIh}} &= \{\O_{1}^{\ell bus},\O_{3}^{\ell bus},\O_{5}^{\ell bus},\O_{7}^{\ell bus},\O_{9}^{\ell bus}\}\ ,\notag \\[1mm]
\overrightarrow{\O_{\rm \sss VIIi}} &= \{\O_{1}^{\ell bub},\O_{5}^{\ell bub}  \}\ .
\notag
\end{align}
The blocks $\hat \gamma_{\rm \sss VIIa} - \hat \gamma_{\rm \sss VIIi}$ are then given by:
\eqa{
\hat \gamma_{\rm\sss VIIa} &=& \frac{\alpha_s}{4\pi}\left(
\begin{array}{ccccc}
 2 & -\frac{1}{4} & 0 & 0 & 0 \\
 16 & -3 & 0 & 0 & 0 \\
 0 & 0 & 1 & \frac{1}{4} & 0 \\
 0 & 0 & -4 & 3 & \frac{1}{4} \\
 0 & 0 & 64 & -16 & -3 \\
\end{array}
\right)+ \frac{\alpha_{\rm em}}{4\pi}\left(
\begin{array}{ccccc}
 -\frac{16}{3} & 0 & 0 & 0 & 0 \\
 0 & -\frac{16}{3} & 0 & 0 & 0 \\
 0 & 0 & \frac{14}{3} & 0 & 0 \\
 0 & 0 & 0 & -\frac{26}{3} & 0 \\
 0 & 0 & 0 & \frac{640}{3} & \frac{14}{3} \\
\end{array}
\right),
\\[2mm]
\hat \gamma_{\rm\sss VIIb} &=& \frac{\alpha_s}{4\pi}\left(
\begin{array}{cc}
 -\frac{1}{2} & 0 \\
 0 & -1 \\
\end{array}
\right)+ \frac{\alpha_{\rm em}}{4\pi}\left(
\begin{array}{cc}
 -\frac{16}{3} & 0 \\
 0 & -\frac{26}{3} \\
\end{array}
\right),
\\[2mm]
\hat \gamma_{\rm\sss VIIc}= \hat \gamma_{\rm\sss VIIh} &=& \frac{\alpha_s}{4\pi}\left(
\begin{array}{ccccc}
 2 & -\frac{1}{4} & 0 & 0 & 0 \\
 16 & -3 & 0 & 0 & 0 \\
 0 & 0 & 1 & \frac{1}{4} & 0 \\
 0 & 0 & -4 & 3 & \frac{1}{4} \\
 0 & 0 & 64 & -16 & -3 \\
\end{array}
\right)+ \frac{\alpha_{\rm em}}{4\pi}\left(
\begin{array}{ccccc}
 -2 & \frac{1}{6} & 0 & 0 & 0 \\
 -\frac{32}{3} & \frac{4}{3} & 0 & 0 & 0 \\
 0 & 0 & -\frac{4}{3} & -\frac{1}{6} & 0 \\
 0 & 0 & \frac{8}{3} & -\frac{8}{3} & -\frac{1}{6} \\
 0 & 0 & -\frac{128}{3} & \frac{32}{3} & \frac{4}{3} \\
\end{array}
\right),
\\[2mm]
\hat \gamma_{\rm\sss VIId}= \hat \gamma_{\rm\sss VIIi} &=& \frac{\alpha_s}{4\pi}\left(
\begin{array}{cc}
 1 & 0 \\
 -\frac{1}{2} & 2 \\
\end{array}
\right)+ \frac{\alpha_{\rm em}}{4\pi}\left(
\begin{array}{cc}
 -\frac{4}{3} & 0 \\
 \frac{1}{3} & -2 \\
\end{array}
\right),
\\[2mm]
\hat \gamma_{\rm\sss VIIe} &=& \frac{\alpha_s}{4\pi}\left(
\begin{array}{ccccc}
 2 & -\frac{1}{4} & 0 & 0 & 0 \\
 16 & -3 & 0 & 0 & 0 \\
 0 & 0 & 1 & \frac{1}{4} & 0 \\
 0 & 0 & -4 & 3 & \frac{1}{4} \\
 0 & 0 & 64 & -16 & -3 \\
\end{array}
\right)+ \frac{\alpha_{\rm em}}{4\pi}\left(
\begin{array}{ccccc}
 -\frac{4}{3} & 0 & 0 & 0 & 0 \\
 0 & -\frac{4}{3} & 0 & 0 & 0 \\
 0 & 0 & -\frac{4}{3} & 0 & 0 \\
 0 & 0 & 0 & -\frac{4}{3} & 0 \\
 0 & 0 & 0 & 0 & -\frac{4}{3} \\
\end{array}
\right),
\\[2mm]
\hat \gamma_{\rm\sss VIIf} &=& \frac{\alpha_s}{4\pi}\left(
\begin{array}{cc}
 -\frac{1}{2} & 0 \\
 0 & -1 \\
\end{array}
\right)+ \frac{\alpha_{\rm em}}{4\pi}\left(
\begin{array}{cc}
 -\frac{4}{3} & 0 \\
 0 & -\frac{4}{3} \\
\end{array}
\right),
\\[2mm]
\hat \gamma_{\rm\sss VIIg} &=& \frac{\alpha_s}{4\pi}\left(
\begin{array}{cc}
 1 & 0 \\
 0 & 2 \\
\end{array}
\right)+ \frac{\alpha_{\rm em}}{4\pi}\left(
\begin{array}{cc}
 -\frac{4}{3} & 0 \\
 0 & -\frac{4}{3} \\
\end{array}
\right).
} 
ADMs for other sets corresponding to primed operators (when existing) or other operators with different flavours,
as specified in Table~\ref{TableList} and Section~\ref{set-up}, are obtained by replication of the appropriate 
matrices given above.

\section{Renormalization-Group Evolution}
\label{RGE}

Given the anomalous dimension matrix and the renormalization group equation (RGE)
\eq{
\frac{d\vec \C}{d\log\mu} = \hat \gamma^T\,\vec \C = 
\frac{\alpha_s}{4\pi} \hat \gamma^{(1,0)T}\,\vec \C 
+ \frac{\alpha_{\rm em}}{4\pi} \hat \gamma^{(0,1)T}\,\vec \C
+ \O(\alpha_s^2, \alpha_{\rm em}^2, \alpha_s \alpha_{\rm em})\ ,
}
the solution for $\C_i(\mu)$ in terms of the initial (matching) conditions $\C_i(\mu_0)$ is expressed in terms of the evolution operator matrix $\hat U(\mu,\mu_0)$,
\eq{
\vec \C(\mu) =\hat  U(\mu,\mu_0) \, \vec \C(\mu_0)
= T\left\lbrace \exp \left[-\frac{1}{2} \int_0^t dt' \,
  \frac{\alpha_s(\mu')}{4\pi} \hat \gamma^{(1,0)T}
+ \frac{\alpha_{\rm em}(\mu')}{4\pi} \hat \gamma^{(0,1)T} \right] \right\rbrace \, \vec \C(\mu_0),
\label{eqRGE}
}
where $t = \ln (\mu_0^2/\mu^2)$ and the $t$-ordered exponential is defined as the Taylor series with each term $t$-ordered, with $t'$ increasing from right to left.
The matrix $ \hat U (\mu,\mu_0)$ can be decomposed as follows:
\begin{equation}
   \hat U (\mu,\mu_0)=
   \hat U_s (\mu,\mu_0)
  + \Delta \hat U_e (\mu,\mu_0),
\label{Utot}
\end{equation}
where $ \hat U_s (\mu,\mu_0)$ is responsible of the evolution in pure QCD, while $\Delta \hat U_e (\mu,\mu_0)$ describes the additional evolution caused by electromagnetic interactions.\footnote{Generalizations to higher orders of these expressions can be found in Ref.~\cite{Buras:1993dy,Huber:2005ig}.}
The leading order result for the pure QCD evolution matrix reads:
\begin{equation}
    \hat U_s(\mu,\mu_0) = 
    \hat V
    \left[ \left(
      \frac{\alpha_s(\mu_0)}{\alpha_s(\mu)}\right)^
    {\frac{\vec \gamma^{(1,0)}}{2\beta^s_0}}
  \right]_D 
  \hat V^{-1},
  \label{eqn:Us}
\end{equation}
where the matrix $\hat V$ diagonalizes $\hat \gamma^{(1,0)T}$,
\eq{
  \hat \gamma_D^{(1,0)}= \hat V^{-1} \, \hat \gamma^{(1,0) T} \,\hat V,
}
and $\vec \gamma^{(1,0)}$ are the diagonal elements of the diagonal matrix $\hat \gamma_D^{(1,0)}$.
The exponent contains the coefficient $\beta_0^s=23/3$ of the leading order QCD beta function $\beta(g_s) = -\beta_0^s \frac{g_s^3}{16 \pi^2}$ with $f=5$ active flavours.
It is convenient to define the parameter $ \eta_s = \eta_s(\mu,\mu_0) \equiv \frac{\alpha_s(\mu_0)}{\alpha_s(\mu)} $ so that we can write the QCD one-loop evolution matrix as
\eq{
  \hat U_s(\mu,\mu_0) = \hat V\, \left[ \eta_s^{\vec a} \right]_D \hat V^{-1},
\label{Uev}
}
with the vector of exponents $\vec a = \vec \gamma^{(1,0)}/(2\beta_0^s)$.

The matrix $\Delta \hat U_e (\mu,\mu_0)$, responsible for the extra evolution in the presence of QED interactions, can be calculated order by order in $\alpha_{\rm em}$; at first order it is given by~\cite{Bellucci:1981bs,Buras:1993dy,Huber:2005ig}:
\eq{
  \Delta \hat U_e (\mu,\mu_0) =
  -\frac{1}{2}
  \int_0^t dt' \,
  \hat U_s(\mu,\mu') \,
  \frac{\alpha_{\rm em} (\mu')}{4\pi} \hat \gamma^{(0,1)T} \,
  \hat U_s(\mu',\mu_0),
  \label{eqn:DeltaUedefinition}
}
where $\mu'$ is such that $t' = \ln (\mu_0^2/\mu' \,^{2})$.
Neglecting the running of $\alpha_{\rm em}$ and employing the leading order expression for $\hat U_s$ in~\eqref{eqn:Us}, the integration in Eq.~\eqref{eqn:DeltaUedefinition} yields
\eq{
  \Delta \hat U_e (\mu,\mu_0) =
  -\frac{\alpha_{\rm em}}{2 \beta_0^s \alpha_s(\mu_0) } \,
   \hat V \hat K(\mu,\mu_0)\hat V^{-1},
  \label{eqn:Ue}
}
where the entries of the matrix $\hat K(\mu,\mu_0)$ are given by:
{\renewcommand{\arraystretch}{3}
\begin{equation}
   K_{ij} (\mu,\mu_0) =
  ( \hat V^{-1} \hat \gamma^{(0,1)T} \hat V)_{ij} \times 
  \begin{cases}
    (\eta_s^{a_j+1}-\eta_s^{a_i})/(a_i-a_j-1)
   & \text{if } a_i-a_j \neq 1,\\
  \eta_s^{a_i}\ln \left( 1/\eta_s \right)
   & \text{if } a_i-a_j = 1.
  \end{cases}
\end{equation}
}

In the rest of this section we provide explicitly the vectors $\vec a$ and the matrices $\hat V$ for each operator Class. 
For convenience, we also provide the complete evolution matrices $ \hat U(\mu,\mu_0)$ in a {\sc mathematica} notebook attached to the arXiv version of this paper~\cite{anc}-- see App.~\ref{notebook}.

\subsection*{Class I : $\mathbf{|\Delta B| = 2}$ operators}

The vector $\vec a$ is given by:
\begin{align}
\vec a_{\rm \sss I}&=\left\{
\frac{6}{23}, \frac{1+\sqrt{241}}{23},\frac{1-\sqrt{241}}{23},
-\frac{24}{23},\frac{3}{23},
\frac{6}{23}, \frac{1+\sqrt{241}}{23},\frac{1-\sqrt{241}}{23}
\right\}.
\end{align}
The matrix $\hat V$ is given by:
\begin{align}
\hat V_{\rm\sss I}&=\left(
\begin{array}{cccccccc}
1 & 0 & 0 & 0 & 0 & 0 & 0 & 0 \\
0 & \frac{-15+\sqrt{241}}{2} & \frac{-15-\sqrt{241}}{2}  & 0 & 0 & 0 & 0 & 0 \\
0 & 1 & 1 & 0 & 0 & 0 & 0 & 0 \\
0 & 0 & 0 & 1 & -\frac{1}{3} & 0 & 0 & 0 \\
0 & 0 & 0 & 0 & 1 & 0 & 0 & 0 \\
0 & 0 & 0 & 0 & 0 & 1 & 0 & 0 \\
0 & 0 & 0 & 0 & 0 & 0& \frac{-15+\sqrt{241}}{2} & \frac{-15-\sqrt{241}}{2} \\
0 & 0 & 0 & 0 & 0 & 0& 1 & 1
\end{array}
\right). 
\end{align}

\subsection*{Class II : $\mathbf{|\Delta B| = 1}$ semileptonic operators}

The exponents $a_i$ are given by:
\eqa{
\vec a_{\rm \sss II}&=\left\{ 0, -\frac{12}{23}, 0, -\frac{12}{23}, \frac{4}{23}\right\}
\ .
}
The  matrix $\hat V$ is simply $\hat V_{\rm\sss II} = \hat {\bf 1}_{5\times 5}$.

\subsection*{Class III : $\mathbf{|\Delta B| = |\Delta C| =1}$  four-quark operators}

The vector $\vec a$ is given by:
\begin{align}
\vec a_{\rm \sss III}&=\left\{
-\frac{24}{23},-\frac{12}{23},\frac{6}{23},\frac{3}{23},\frac{-17-\sqrt{241}}{23},
-\frac{24}{23},\frac{1+\sqrt{241}}{23},\frac{1-\sqrt{241}}{23},
\frac{3}{23},\frac{\sqrt{241}-17}{23}
\right\}. 
\end{align}
As in Eqs.~(\ref{OIII}),(\ref{gammaIII}) we decompose the matrix $\hat V$ into two sub-blocks:
\begin{equation}
\hat V_{\rm \sss III} =
\begin{pmatrix}
\hat V^{1-4}_{\rm \sss III} & 0 \\
0 & \hat V^{5-10}_{\rm \sss III} \\
\end{pmatrix}
\ , 
\end{equation}
with
\begin{equation}
\hat V^{1-4}_{\rm \sss III} =
\left(
\begin{array}{cccc}
 -\frac{8}{3} & \frac{4}{3} & -\frac{8}{3} & \frac{64}{3} \\
 -16 & -4 & -4 & -16   \\
 \frac{1}{6} & -\frac{1}{3} & \frac{2}{3} & -\frac{4}{3} \\
 1 & 1 & 1 & 1 
\end{array}
\right)
\ ,
\end{equation}

{\renewcommand{\arraystretch}{1.5}
\begin{equation}
\hat V^{5-10}_{\rm \sss III} =
\left(
\begin{array}{cccccc}
 -\frac{53}{3}-\sqrt{241} & -64 & \frac{86}{15}-\frac{2 \sqrt{241}}{5} & \frac{2(43+3 \sqrt{241})}{15} & 0 & -\frac{53}{3}+\sqrt{241} \\
 -16 & 0 & -16 & -16 & -64 & -16 \\
 \frac{79}{4}+\frac{11 \sqrt{241}}{12} & 16 & \frac{-207+7 \sqrt{241}}{30} & \frac{-207-7\sqrt{241}}{30} & 0 & \frac{79}{4}-\frac{11 \sqrt{241}}{12} \\
 27+\sqrt{241} & 0 & \frac{2(51-\sqrt{241})}{5} & \frac{2(51+\sqrt{241})}{5}  & 16 & 27-\sqrt{241} \\
 \frac{53+3 \sqrt{241}}{48} & 1 & \frac{-43+3 \sqrt{241}}{120} & \frac{-43-3\sqrt{241}}{120} & 0 & \frac{53-3 \sqrt{241}}{48} \\
 1 & 0 & 1 & 1 & 1 & 1 \\
\end{array}
\right).
\end{equation}
}

\subsection*{Class IV : $\mathbf{|\Delta B| = 1,\ |\Delta S| = 2}$  operators}

The exponents $a_i$ are given by:
\eqa{
\vec a_{\rm\sss IV}&=\left\{-\frac{24}{23},\frac{1+\sqrt{241}}{23},\frac{1-\sqrt{241}}{23},\frac{6}{23},\frac{3}{23}\right\}
\ .
}
The matrix $\hat V$ is given by:
\eq{
\hat V_{\rm\sss IV}=\left(
\begin{array}{ccccc}
 0 & 0 & 0 & -4 & -96 \\
 0 & 0 & 0 & 1 & 6 \\
 -64 & -16 & -16 & 0 & -64 \\
 16 & \frac{4(36+\sqrt{241})}{5}  & \frac{4(36-\sqrt{241})}{5}  & 0 & 16 \\
 1 & 1 & 1 & 0 & 1 \\
\end{array}
\right)
\ . 
}

\subsection*{Class V : $\mathbf{|\Delta B| = 1,\ |\Delta C| = 0}$  operators}

The corresponding vector $\vec a$ and matrix $\hat V$ in this class are 57-dimensional, and thus it is not practical to present them explicitly here. 
In addition, the diagonalization of the ADM block $\hat \gamma_{\rm\sss V}$ cannot be carried out
analytically, and therefore the expressions are necessarily numerical with finite precision. 
The complete numerical expression for the $57\times 57$ evolution matrix $\hat U_{\rm \sss V}(\mu,\mu_0)$ can be found as {\sc mathematica} notebook in~\cite{anc}.

\subsection*{Class VI : Lepton Number Violating operators}
The vectors $\vec a$ and the matrices $\hat V$ for classes VIa-c are
\begin{align}
\vec{a}_{\rm\sss VIa} & = 
\left\lbrace 0 , -\frac{12}{23} , \frac{4}{23}, 0, -\frac{12}{23} \right\rbrace\, ,\\
\hat{V}_{\rm\sss VIa} & = \hat {\bf 1}_{5\times 5}\, ,\\
\vec{a}_{\rm\sss VIb} & = \vec{a}_{\rm\sss VIc} = 
\left\lbrace -\frac{12}{23} , \frac{4}{23}, -\frac{12}{23} \right\rbrace \, ,\\[1mm]
\hat{V}_{\rm\sss VIb} & = \hat{V}_{\rm\sss VIc} = \hat {\bf 1}_{3\times 3} \, .
\end{align}
\subsection*{Class VII : Baryon Number Violating operators}

For Class VIIa, VIIc, VIIe and VIIh the vectors $\vec a$ and the matrices $\hat V$ are given by:
\eq{
\vec a_{\rm\sss VIIa}=\vec a_{\rm\sss VIIc}=\vec a_{\rm\sss VIIe}={ \vec a_{\rm\sss VIIh}}=\left\{
-\frac{3}{23},\frac{3}{46},-\frac{3}{23},\frac{3}{23},\frac{3}{46}
\right\}
\ , \quad
}
and 
\eq{
\hat V_{\rm\sss VIIa}=\hat V_{\rm\sss VIIc}=\hat V_{\rm\sss VIIe}={ \hat V_{\rm\sss VIIh}}=\left(
\begin{array}{ccccc}
 -4 & -16 & 0 & 0 & 0 \\
 1 & 1 & 0 & 0 & 0 \\
 0 & 0 & -16 & -16 & -64 \\
 0 & 0 & 4 & 20 & 16 \\
 0 & 0 & 1 & 1 & 1 \\
\end{array}
\right).
}

\noindent For Class VIId and VIIi the vector $\vec a$ and the matrix $\hat V$ are given by:
\eq{
\vec a_{\rm\sss VIId}={ \vec a_{\rm\sss VIIi}}= \left\{
\frac3{23},\frac3{46}
\right\}
\ ,
}
\eq{
\hat V_{\rm\sss VIId}=
{ \hat V_{\rm\sss VIIi}} =
\left(
\begin{array}{cc}
 -\frac12 & 1 \\
 1 & 0  \\
\end{array}
\right)
\ .
}

\noindent For the Classes VIIb, VIIf and VIIg, the anomalous dimensions are diagonal and therefore
\eq{
\hat V_{\rm\sss VIIb} = 
\hat V_{\rm\sss VIIf} = 
\hat V_{\rm\sss VIIg} = 
\hat {\bf 1}_{2\times 2}\ ,
}
and the vectors $\vec a$ are given by
\begin{align}
\vec a_{\rm\sss VIIb} &= \vec a_{\rm\sss VIIf} = \left\{ -\frac3{92}, -\frac3{46} \right\}
\ ,
&
\vec a_{\rm\sss VIIg} = \left\{ \frac3{46}, \frac3{23} \right\}
\ .
\end{align}

\section{Numerical Example: Class-V \emph{Spectra}}
\label{sec:spectra}

\begin{figure}
  \centering
  \subfloat[][Contribution to the Wilson coefficients at the scale $\mu_b$ assuming $\C_5^{sbbb}(\mu_W)=1$.]{\includegraphics[width=0.99\textwidth]{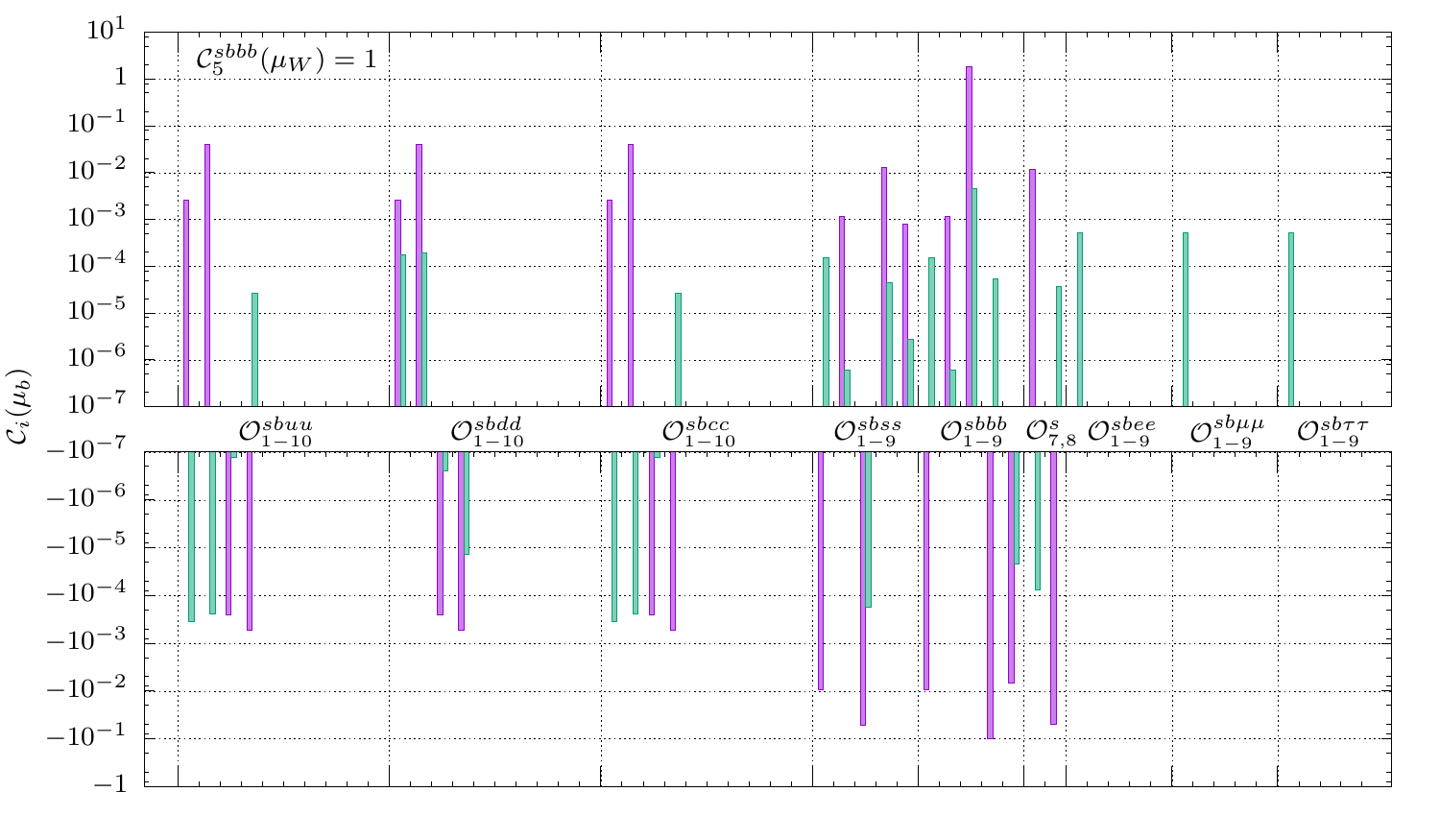}\label{fig:Q5b}}\\
  \subfloat[][Contributions to $\C_{7\gamma}^s(\mu_b)$ for a matching condition $\C_j(\mu_W)=1$.]{\includegraphics[width=0.99\textwidth]{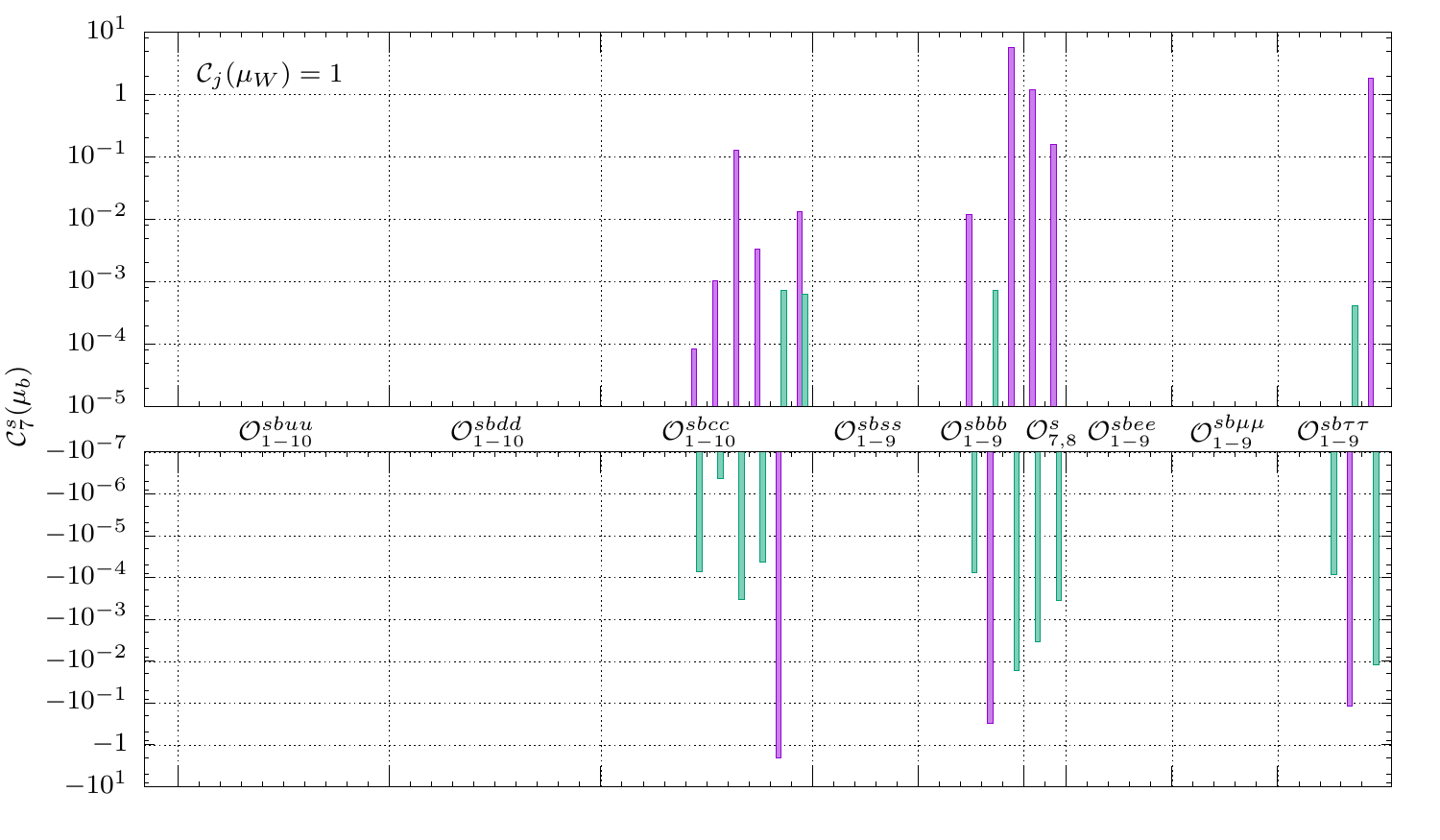}\label{fig:Q7s}}
  \caption{Examples of \emph{spectra} of Class V.
   Purple and green bars correspond the QCD and QED renormalization-group evolution given by the matrices $\hat U_s$ and $\Delta \hat U_e$, respectively.}
  \label{fig:example}
\end{figure}

\begin{figure}
  \centering
  \includegraphics[width=0.99\textwidth]{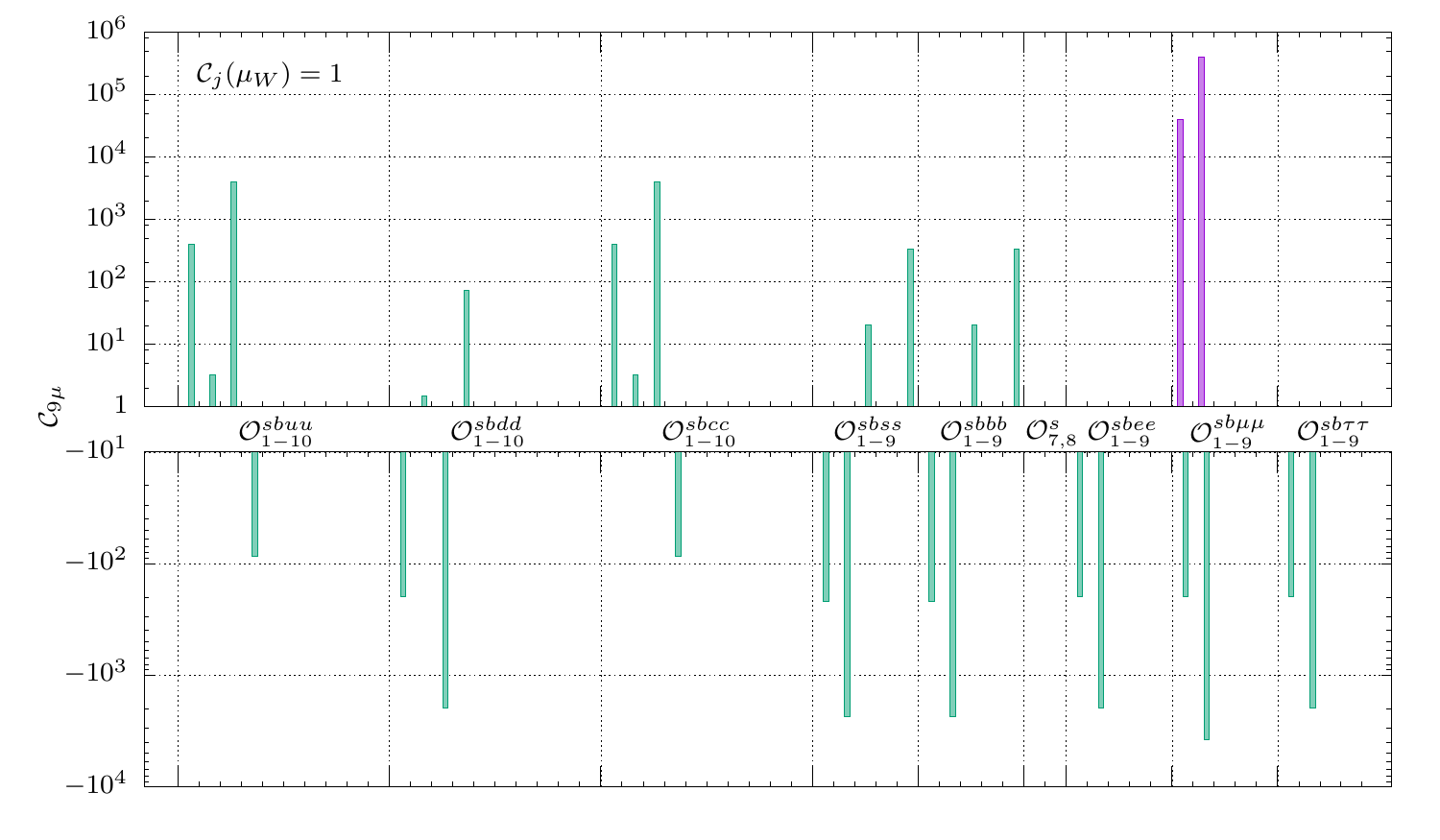}
  \caption{Contributions to $\C_{9\mu}$ as defined in~\eqref{eqn:C9mu} assuming  $\C_j(\mu_W)=1$ for each operator $\O_j$ in Class V.}
  \label{fig:Q9mu}
\end{figure}

As mentioned in the previous section, the number of operators in Class V is too large to present here explicitly
the evolution matrix $\hat U_V$ (see App.~\ref{notebook}).
Nevertheless, we would like to discuss a simple way to visualize the matrix $\hat U_V$ by making use of bar plots, as those presented in Fig.~\ref{fig:example}.

The solution of the RGE~\eqref{eqRGE} can be written in components as
\begin{equation}
  \C_i(\mu_b) = \sum_j 
  \hat U_{ij} (\mu_b,\mu_W) \, \C_j(\mu_W)
  = \sum_j 
  \left[ \hat U_s(\mu_b,\mu_W)+\Delta \hat U_e(\mu_b,\mu_W) \right]_{ij} \, \C_j(\mu_W).
  \label{eqn:RGEcomponent}
\end{equation}
The values of the Wilson coefficients can be displayed in a bar plot providing a sort of \emph{spectrum} of Class V.
We distinguish two types of plots: we can show all $\C_i(\mu_b)$, for a given set of matching conditions $\C_j(\mu_W)$,
or for a fixed $i$ we can show all single terms appearing in the $j$-summation in Eq.~\eqref{eqn:RGEcomponent} stemming
from each $\C_j(\mu_W)$. These two plots can be employed to convey different types of information:

\noindent $\blacktriangleright$ \textbf{\boldmath $\C_i(\mu_b)$-spectrum}: it shows the value of all Wilson coefficients at the scale $\mu_b$, $\C_i(\mu_b)$, for a given set of matching conditions $\C_j(\mu_W)$.

A simple example is given in Fig.~\ref{fig:Q5b}.
Each operator $\O_i$ in Class V corresponds to a bin on the $x$-axis; its Wilson coefficient at the scale $\mu_b$
is represented by a bar (with positive or negative value).
As matching condition we simply set $\C_5^{sbbb}(\mu_W) = 1$ and all others equal to zero. 
Purple and green bars correspond to the QCD and QED contributions given by the matrices $\hat U_s$ and $\Delta \hat U_e$, respectively. The two scales are chosen to be $\mu_W = M_Z$ and $\mu_b=m_b$. 

In general, more than one $\C_j(\mu_W)$ is different from zero, so that the sum over $j$ must be taken in Eq.~\eqref{eqn:RGEcomponent}. 
For instance, once a specific new physics scenario is considered and the whole set of matching conditions is known,
the $\C_i(\mu_b)$-spectrum gives an overall view of the sizes of the Wilson coefficients at the scale $\mu_b$.

\bigskip

\noindent $\blacktriangleright$ \textbf{\boldmath $\C_j(\mu_W)$-spectrum}: it shows, for a fixed $i$,
each partial contribution to $\C_i(\mu_b)$ in the sum~\eqref{eqn:RGEcomponent}.

Fig.~\ref{fig:Q7s} shows each partial contribution to $\C_{7\gamma}^s(\mu_b)$ for an initial condition $\C_j(\mu_W) = 1$ (for all $j$); operator names are on the $x$-axis.
We note that the bars can be viewed also as the value of $\C_i(\mu_b)$ if only the corresponding $\C_j(\mu_W)$ is set
to be non-zero at the scale $\mu_W$. From this perspective, suppose that $|\C_i(\mu_b)| < k$, then $k$ times the inverse of
the bar size can be regarded as the corresponding constraint on $\C_j(\mu_W)$. In our case we could read for
example $|\C_9^{sbbb}(\mu_W)| \lesssim k/5$ or $|\C_5^{sbcc}(\mu_W)| \lesssim k/10^{-4}$ etc. It is understood
that this rough estimate holds under the assumption that only one Wilson coefficient is different from zero at the
scale $\mu_W$.

\bigskip

The same kind of spectra can be drawn for linear combinations of Wilson coefficients; for example Fig.~\ref{fig:Q9mu} shows the $\C_{9\mu}$-spectrum of the SM-like operator $\C_{9\mu}$ defined as
\begin{equation}
  \frac{\alpha_{\rm em}}{4\pi} \, V_{tb}V_{ts}^* \, \C_{9\mu} (\mu_b) =
  \C^{sb\mu\mu}_1(\mu_b)+10 \, \C^{sb\mu\mu}_3(\mu_b),
  \label{eqn:C9mu}
\end{equation}
with $\C_j(\mu_W)=1$ for all $j$.

\section{Conclusion}
\label{conclusion}

General analyses of $B$-physics processes beyond the SM require control of the renormalization-group evolution
below the electroweak scale. This evolution is well known for the dimension-six operators in the WET that have non-negligible matching conditions in the SM. However, in a general New Physics model, many other operators
may receive relevant matching conditions. The first step is to write down the most general set of dimension-six
operators in the WET. We have built a complete, minimal and suitable basis of operators relevant for $B$-physics.
This basis is presented in Section~\ref{set-up}.   

We have also calculated and collected the complete set of one-loop anomalous dimensions of these operators.
The anomalous dimension matrices for each operator class can be found in Section~\ref{ADM}.
The evolution equation for the Wilson coefficients necessary to
evaluate the coefficients at the $B$ physics scale in terms of the matching conditions at the EW or the New Physics scale,
with resummation of QCD and QED leading logarithms is given in Eqs.~\eqref{Utot},\eqref{Uev}~and~\eqref{eqn:Ue}.
The explicit results for the different blocks
corresponding to the different classes of operators (see Section~\ref{set-up}) are also given in Section~\ref{RGE}.
The evolution matrices are given for convenience in electronic format as a {\sc mathematica} package attached to this
paper \cite{anc}, and discussed in App.~\ref{notebook}.

The results of this paper will be useful in any attempt to automatize completely general analyses of physics beyond the SM
which take into account consistently experimental constraints from $B$-physics. These results have already been incorporated into the modular program {\tt DsixTools}~\cite{Celis:2017hod} and recently also in \textsf{\emph{wilson}}~\cite{Aebischer:2018bkb}.

\section*{Acknowledgements}
We thank Thomas Mannel and Avelino Vicente for suggestions.
We also thank Mikolaj Misiak for pointing out to us a few missing lepton-number-violating operators.
M.F.\ thanks H.\ Patel for help with {\tt Package-X}~\cite{Patel:2015tea}. 
This work is supported by the Swiss National Science Foundation.
J.V. acknowledges additional funding from Explora project FPA2014-61478-EXP.

\appendix

\section{Complete Numerical Results for the RG Evolution Matrices}
\label{notebook}

From the results presented in Section~\ref{RGE} one can easily construct all the matrices $\hat U(\eta_s)$
needed for the evolution of all Wilson coefficients. In the case of $\hat U_{\rm \sss V}(\eta_s)$,
we have not presented the explicit expressions for the 57-dimensional vector $\vec a_{\rm\sss V}$
and rotation matrix $V_{\rm\sss V}$, but they can be obtained by diagonalization of the ADM given in
Eq.~(\ref{eqn:ADMclassV}).

For convenience, we provide a {\sc mathematica} package~\cite{anc}
called {\tt EvolutionMatrices.m}, which contains all the matrices $\hat U_J(\eta_s)$ for
$J = {\rm I, II,\dots,VIIi}$, as a function of the coupling ratio $\eta_s\equiv \alpha_s(\mu_0)/\alpha_s(\mu)$ and
the QED fine-structure constant $\alpha_{\rm em}$.
After correctly specifying the path and evaluating the package

\medskip

\noindent {\tt <<``EvolutionMatrices.m''}

\medskip

\noindent the variables {\tt Us[I]}, {\tt Us[II]}, etc. contain the QCD evolution matrices $\hat U_s$ corresponding to each
class of operators, and the variables , {\tt Ue[I]}, {\tt Ue[II]}, etc. contain the corresponding QED contributions
$\Delta \hat U_e$. For example, the complete evolution for Class~V operators to first order in QED is obtained doing:\\[3mm]
\noindent {\tt <<``EvolutionMatrices.m''}\\
\noindent {\tt UclassV = Us[V] + Ue[V];}\\[3mm]
and similarly for the other operator Classes.

\section{Fierz Identities for Four Quark Operators}
\label{fierz}
In this Appendix we give the (four-dimensional) Fierz identities that allow to remove the redundant color-octet four-quark operators
in Classes~IV and~V. The discussion is framed in the context of Class~V operators, but the case of Class~IV
is completely analogous as for Class~V $sbss$ operators.

It is convenient, also for the following comparison with previously published results, to introduce the following
\emph{Fierz} basis of four-quark operators. For $q=u,d,c,b$ we define:
\begin{align}
  \fop_1^q &=
  \left( \bar{s} \gamma_\mu P_L b\right) \left( \bar{q} \gamma^\mu P_L q\right), &
  \fop_2^q &=
  \left(\bar{s}_\alpha \gamma_\mu P_L b_\beta\right) \left( \bar{q}_\beta \gamma^\mu P_L q_\alpha\right), \notag \\
  \fop_3^q &=
  \left( \bar{s} \gamma_\mu P_L b\right) \left( \bar{q} \gamma^\mu P_R q\right), &
  \fop_4^q &=
  \left(\bar{s}_\alpha \gamma_\mu P_L b_\beta\right) \left( \bar{q}_\beta \gamma^\mu P_R q_\alpha\right), \notag \\
  \fop_{5}^q &= (\bar{s}P_R b) (\bar{q} P_R q)\,, &
  \fop_{6}^q &= (\bar{s}_\alpha P_R b_\beta) (\bar{q}_\beta P_R q_\alpha)\,, \notag \\
  \fop_{7}^q &= (\bar{s}P_R b) (\bar{q} P_L q)\,, &
  \fop_{8}^q &= (\bar{s}_\alpha P_R b_\beta) (\bar{q}_\beta P_L q_\alpha)\,, \notag\\
  \fop_{9}^q &= (\bar{s} \sigma^{\mu\nu} P_R b) (\bar{q} \sigma_{\mu\nu} P_R q)\,, &
  \fop_{10}^q &= (\bar{s}_\alpha \, \sigma^{\mu\nu} P_R b_\beta) (\bar{q}_\beta \, \sigma_{\mu\nu} P_R q_\alpha)\,;
  \label{eqn:fopq}
\end{align}
while for $q=s$:
\begin{align}
  \fop_1^s &=
  \left( \bar{s} \gamma_\mu P_L b\right) \left( \bar{s} \gamma^\mu P_L s\right), &
  \fop_2^s &=
  \left(\bar{s}_\alpha \gamma_\mu P_L b_\beta\right) \left( \bar{s}_\beta \gamma^\mu P_L s_\alpha\right), \notag \\
  \fop_3^s &=
  \left( \bar{s} \gamma_\mu P_L b\right) \left( \bar{s} \gamma^\mu P_R s\right), &
  \fop_4^s &=
  \left(\bar{s}_\alpha \gamma_\mu P_L b_\beta\right) \left( \bar{s}_\beta \gamma^\mu P_R s_\alpha\right), \notag \\
  \fop_{5}^s &= (\bar{s}P_L b) (\bar{s} P_L s)\,, &
  \fop_{6}^s &= (\bar{s}_\alpha P_L b_\beta) (\bar{s}_\beta P_L s_\alpha)\,, \notag \\
  \fop_{7}^s &= (\bar{s}P_L b) (\bar{s} P_R s)\,, &
  \fop_{8}^s &= (\bar{s}_\alpha P_L b_\beta) (\bar{s}_\beta P_R s_\alpha)\,, \notag\\
  \fop_{9}^s &= (\bar{s} \sigma^{\mu\nu} P_L b) (\bar{s} \sigma_{\mu\nu} P_L s)\,, &
  \fop_{10}^s &= (\bar{s}_\alpha \, \sigma^{\mu\nu} P_L b_\beta) (\bar{s}_\beta \, \sigma_{\mu\nu} P_L s_\alpha)\,.
  \label{eqn:fops}
\end{align}
The analogous set of primed operators with opposite chirality is obtained interchanging $P_L \leftrightarrow P_R$ everywhere.
For $q=s,b$ not all operators are independent and Fierz identities in $D=4$ allow to remove half of them. In this work, we choose to express the even operators in terms of the odd ones via the identities (with anticommuting fermion fields):
\begin{align}
  \fop_2^{s,b} &= \fop_1^{s,b}, \notag &
  \fop_4^{s,b} &= -2\fop_7^{s,b},\notag  \\
  \fop_6^{s,b} &= -\frac{1}{2}\fop_5^{s,b} -\frac{1}{8}\fop_9^{s,b},\notag &
  \fop_8^{s,b} &= -\frac{1}{2}\fop_3^{s,b},\notag \\
  \fop_{10}^{s,b} &= -6\fop_5^{s,b} +\frac{1}{2}\fop_9^{s,b}.
  \label{eqn:fierzids}
\end{align}
Note that with the operator definitions given in Eqs.~\eqref{eqn:fopq} and~\eqref{eqn:fops}, primed and unprimed
operators do not mix, which is the main reason for the different definition in $q=s$ operators.
The reason for choosing to eliminate the color-octet operators (the ones with even indices), is that one-loop
closed penguins involving $\O^{sbss}$ or $\O^{sbbb}$ will not appear.

Using the identities (\ref{eqn:idthreegammas}-\ref{eqn:idsigma}) and the relation among matrices of the fundamental representation of $SU(N)$,
\begin{equation}
  T^A_{ij} T^A_{kl} = \frac{1}{2}\delta_{il}\delta_{kj}-\frac{1}{2N}\delta_{ij}\delta_{kl},
\end{equation}
the $\fop$ operators can be expressed in terms of the four-quark operators of Class V in Eqs.~\eqref{sbqq} and~\eqref{sbss} by means of the following linear transformation:
\begin{equation}
  \fop_i^q =  R_{ij}\, \O^{sbqq}_j, 
   \qquad 
   q=u,d,s,c,b,
   \label{eqn:fierztonew}
\end{equation}
where $\hat R=\mathop{\mathrm{diag}}(\hat R_A, \hat R_B)$ is a block diagonal matrix where the $4\times4$ sub-block $\hat R_A$ maps $\O_{1-4}$ into $\fop_{1-4}^q$, and the $6\times6$ sub-block $\hat R_B$ maps $\O_{5-10}$ into $\fop_{5-10}^q$; their explicit expressions are
{
\renewcommand{\arraystretch}{1.5}
\begin{align}
  \hat R_A & =
  \begin{pmatrix}
   -\frac{1}{3} & 0 & \frac{1}{12} & 0 \\
   -\frac{1}{3 N} & -\frac{2}{3} & \frac{1}{12 N} & \frac{1}{6} \\
   \frac{4}{3} & 0 & -\frac{1}{12} & 0 \\
   \frac{4}{3 N} & \frac{8}{3} & -\frac{1}{12 N} & -\frac{1}{6} \\
  \end{pmatrix}, &
  \hat R_B &=
  \begin{pmatrix}
   -\frac{1}{3} & 0 & \frac{1}{3} & 0 & \frac{1}{48} & 0 \\
   -\frac{1}{3 N} & -\frac{2}{3} & \frac{1}{3 N} & \frac{2}{3} &
   \frac{1}{48 N} & \frac{1}{24} \\
   \frac{4}{3} & 0 & -\frac{1}{3} & 0 & -\frac{1}{48} & 0 \\
   \frac{4}{3 N} & \frac{8}{3} & -\frac{1}{3 N} & -\frac{2}{3} &
   -\frac{1}{48 N} & -\frac{1}{24} \\
   0 & 0 & 1 & 0 & 0 & 0 \\
   0 & 0 & \frac{1}{N} & 2 & 0 & 0 \\  \end{pmatrix}.
\end{align}}
The same transformation applies to primed operators.
Eq.~\eqref{eqn:fierztonew} allows us to obtain the Fierz identities for the operators $\O^{sbbb}$ and $\O^{sbss}$
in Eqs.~(\ref{sbqq}) and (\ref{sbss}):
\begin{align}
  \O_2&=-\frac{1}{3}\O_1+\frac{1}{24}\O_3-\frac{4}{3}\O_5+\frac{1}{3}O_7+\frac{1}{48}\O_9\ ,
  \notag \\[1mm]
  \O_4&=-\frac{8}{3}\O_1+\frac{1}{2}\O_3-\frac{16}{3}\O_5+\frac{4}{3}\O_7+\frac{1}{12}\O_9\ ,
  \notag \\[1mm]
  \O_6&=-\frac{1}{3}\O_1+\frac{1}{48}\O_3-\frac{1}{12}\O_5-\frac{7}{48}\O_7-\frac{1}{192}O_9\ ,
  \notag \\[1mm]
  \O_8&=\O_5-\frac{11}{12}\O_7-\frac{1}{16}\O_9\ ,
  \notag \\[1mm]
  \O_{10}&=-\frac{16}{3}\O_1+\frac{1}{3}\O_3-\frac{32}{3}\O_5+\frac{8}{3}\O_7+\frac{1}{2}\O_9\ .
\end{align}
The same 4D identities hold for the primed operators, for Class~IV operators (\ref{sbsd}), and for the corresponding
operators with $s\leftrightarrow d$. These identities are useful in the calculation of the one-loop anomalous dimensions,
and set a reference for the subsequent definition of Fierz evanescent operators necessary for fixing the scheme in higher order
calculations~\cite{Buras:2000if}.

\section{Semileptonic Operators : Traditional Basis}
\label{SemileptonicBasis}

In this Appendix we provide the transformation rules to translate the Wilson coefficients of semileptonic operators
between our basis and a more ``traditional'' one, e.g. Refs.~\cite{1512.02830,1510.04239,1212.2321}.

\bigskip

\noindent $\blacktriangleright$ {\bf Class~II operators:}\\[2mm]
We consider the basis in Ref.~\cite{1512.02830} for $|\Delta B| =|\Delta C| =1$ operators.
In this case the operators are equivalent to ours, with a redefinition of primed and unprimed operators necessary to
block-diagonalize the ADM. The dictionary is given by:
\eq{
C_V = C_1^{cb\ell\ell'}\ ,\quad
C'_V = C_{1'}^{cb\ell\ell'}\ ,\quad
C_S = C_{5'}^{cb\ell\ell'}\ ,\quad
C'_S = C_{5}^{cb\ell\ell'}\ ,\quad
C_T = C_{7'}^{cb\ell\ell'}\ .
}
The same relations hold for $b\to u\ell\nu_{\ell'}$.

\bigskip

\noindent $\blacktriangleright$ {\bf Semileptonic Class~V operators:}\\[2mm]
The translation in this case requires a bit of work. 
We start with the ``Fierz'' basis for semileptonic operators:
\begin{align}
  \fop^{\ell\ell^\prime}_9 &= %
  (\bar{s}\,\gamma_\mu P_L b) \; (\bar{\ell} \gamma^\mu \ell^\prime)\,, &  
  \fop^{\ell\ell^\prime}_{10} &= %
  (\bar{s} \gamma_\mu P_L b) \;(\bar{\ell} \gamma^\mu \gamma_5 \ell^\prime)\,,\notag \\
  \fop^{\ell\ell^\prime}_S &= (\bar{s} P_R b) \; (\bar{\ell} \ell^\prime)\,,&  
  \fop_P^{\ell \ell'} &= (\bar{s} P_R b) \; (\bar{\ell} \gamma_5 \ell^\prime)\,,\notag\\
  \fop^{\ell\ell^\prime}_T & =(\bar{s} \sigma^{\mu\nu} b) \; (\bar{\ell} \sigma_{\mu\nu} \ell^\prime)\,,&
  \fop^{\ell\ell^\prime}_{T5} & =(\bar{s} \sigma^{\mu\nu} b) \; (\bar{\ell} \sigma_{\mu\nu} \gamma_5 \ell^\prime)\, ,
  \label{eqn:fierzbasislep}
\end{align}
plus the four primed operators $\fop^{\ell \ell'}_{9',10',S',P'}$
obtained from the unprimed by interchanging $P_L \leftrightarrow P_R$. 
The operators $\fop^{\ell \ell'}$ are given in terms of Class-V semileptonic operators in Eq.~\eqref{sbll} by
\begin{equation}
  \fop_i^{\ell \ell'} =
  R^\ell_{ij}\, \O_{j}^{sb\ell \ell'}\ ,
\end{equation}
where we have combined the operators in the following way:
\begin{align}
  \overrightarrow{\fop}^{\ell \ell'} &=
  \{\fop^{\ell\ell^\prime}_9 ,\fop^{\ell\ell^\prime}_{10},
  \fop^{\ell\ell^\prime}_S ,\fop^{\ell\ell^\prime}_P ,
  \fop^{\ell\ell^\prime}_T ,\fop^{\ell\ell^\prime}_{T5} ,
  \fop^{\ell\ell^\prime}_{9'} ,\fop^{\ell\ell^\prime}_{10'},
  \fop^{\ell\ell^\prime}_{S'} ,\fop^{\ell\ell^\prime}_{P'}\}\ ,\\[2mm]
  \overrightarrow{\O}^{sb\ell \ell'} &=
  \{\O^{sb\ell \ell'}_{1-9},\O^{' \, sb \ell \ell'}_{1-9}\}\ . 
\end{align}
The explicit expression of the $10 \times 10$ matrix $\hat R_\ell$ is
\begin{equation}
 \hat R_\ell =
 \begin{pmatrix}
 1 & 0 & 0 & 0 & 0 & 0 & 0 & 0 & 0 & 0 \\
 \frac{5}{3} & -\frac{1}{6} & 0 & 0 & 0 & 0 & 0 & 0 & 0 & 0 \\
 0 & 0 & 1 & 0 & 0 & 0 & 0 & 0 & 0 & 0 \\
 0 & 0 & -\frac{5}{3} & \frac{2}{3} & \frac{1}{24} & 0 & 0 & 0 & 0 & 0 \\
 0 & 0 & 0 & 1 & 0 & 0 & 0 & 0 & 1 & 0 \\
 0 & 0 & 0 & 1 & 0 & 0 & 0 & 0 & -1 & 0 \\
 0 & 0 & 0 & 0 & 0 & 1 & 0 & 0 & 0 & 0 \\
 0 & 0 & 0 & 0 & 0 & -\frac{5}{3} & \frac{1}{6} & 0 & 0 & 0 \\
 0 & 0 & 0 & 0 & 0 & 0 & 0 & 1 & 0 & 0 \\
 0 & 0 & 0 & 0 & 0 & 0 & 0 & \frac{5}{3} & -\frac{2}{3} & -\frac{1}{24} \\ 
 \end{pmatrix}.
\end{equation}

\bigskip

\noindent We define the ``traditional'' basis of operators and Wilson coefficients by the Lagrangian:
\eqa{
\L_{\rm s.l.}^{b\to s} &=& \frac{4G_F}{\sqrt{2}} V_{tb}V^\star_{ts}
\bigg\{  C_7\,O_7 + C_8\,O_8 + C_{7'}\,O_{7'} + C_8\,O_8 + C_{8'} O_{8'}
\nonumber\\[2mm]
&& 
+ \sum_{\ell=e,\mu,\tau} \Big[ C_{9}^\ell\,O_{9}^\ell + C_{10}^\ell\,O_{10}^\ell
+ C_{S}^\ell\,O_{S}^\ell + C_{P}^\ell\,O_{P}^\ell + C_{T}^\ell\,O_{T}^\ell
+ C_{T5}^\ell\,O_{T5}^\ell
\nonumber\\
&&\hspace{2cm}
+ C_{9'}^\ell\,O_{9'}^\ell + C_{10'}^\ell\,O_{10'}^\ell + C_{S'}^\ell\,O_{S'}^\ell + C_{P'}^\ell\,O_{P'}^\ell  \Big]
\bigg\}\ ,
}
where the different operators are related to our operators by:
\eq{
O_{i}^\ell = \frac{\alpha_{\rm em}}{4\pi} \fop_i^{\ell\ell}\ ,\quad
O_{7^{(\prime)}} = \frac{\alpha_s}{4\pi} \O^s_{7^{(\prime)}\gamma} \ ,\quad
O_{8^{(\prime)}} = \frac{\alpha_s}{4\pi} \O^s_{8^{(\prime)}g} \ .
}
These definitions are consistent with Refs.~\cite{1510.04239,1212.2321}, but not with Ref.~\cite{1512.02830} where
the CKM elements are not factored out. Thus it is important to have this in mind when using the matching conditions
in Ref.~\cite{1512.02830}. These definitions are also consistent with the usual values quoted for the SM Wilson
coefficients: $C_7^{\rm SM}(m_b) \simeq -0.3$, $C_8^{\rm SM}(m_b) \simeq -0.17$ and
$C_9^{\ell\ {\rm SM}}(m_b) \simeq  -C_{10}^{\ell\ {\rm SM}}(m_b) \simeq 4$.

It follows that the Wilson coefficients $C_i$ in this ``traditional'' basis are related to the Wilson coefficients
of Class-V operators in Eqs.~(\ref{sbM}) and \eqref{sbll} by
\begin{align}
  \frac{\alpha_s}{4\pi} V_{tb}V^\star_{ts}\ C_7 &= \C_{7\gamma}^{s}\ , &
  \frac{\alpha_s}{4\pi} V_{tb}V^\star_{ts}\ C_{7'}^\ell &= \C_{7'\gamma}^{s}\ , \notag \\[2mm]
  \frac{\alpha_s}{4\pi} V_{tb}V^\star_{ts}\ C_8 &= \C_{8g}^{s}\ , &
  \frac{\alpha_s}{4\pi} V_{tb}V^\star_{ts}\ C_{8'}^\ell &= \C_{8'g}^{s}\ , \notag \\[2mm]
  \frac{\alpha_{\rm em}}{4\pi} V_{tb}V^\star_{ts}\ C_9^\ell &= \C_1^{sb\ell\ell} + 10 \, \C_3^{sb\ell\ell}\ , &
  \frac{\alpha_{\rm em}}{4\pi} V_{tb}V^\star_{ts}\ C_{9'}^\ell &= \C_{1'}^{sb\ell\ell} + 10 \, \C_{3'}^{sb\ell\ell}\ , \notag \\[2mm]
  \frac{\alpha_{\rm em}}{4\pi} V_{tb}V^\star_{ts}\ C_{10}^\ell &= -6 \, \C_3^{sb\ell\ell}\ , &
  \frac{\alpha_{\rm em}}{4\pi} V_{tb}V^\star_{ts}\ C_{10'}^\ell &= 6 \, \C_{3'}^{sb\ell\ell}\ ,\notag  \\[2mm]
  \frac{\alpha_{\rm em}}{4\pi} V_{tb}V^\star_{ts}\ C_S^\ell &= \C_5^{sb\ell\ell} + 40 \, \C_9^{sb\ell\ell}\ , &
  \frac{\alpha_{\rm em}}{4\pi} V_{tb}V^\star_{ts}\ C_{S'}^\ell &= \C_{5'}^{sb\ell\ell} + 40 \, \C_{9'}^{sb\ell\ell}\ ,\notag  \\[2mm]
  \frac{\alpha_{\rm em}}{4\pi} V_{tb}V^\star_{ts}\ C_{P}^\ell &= 24 \, \C_3^{sb\ell\ell}\ , &
  \frac{\alpha_{\rm em}}{4\pi} V_{tb}V^\star_{ts}\ C_{P'}^\ell &= -24 \, \C_{3'}^{sb\ell\ell}\ ,\notag  \\[2mm]
  \frac{\alpha_{\rm em}}{4\pi} V_{tb}V^\star_{ts}\ C_T^\ell &= \frac{1}{2} \, \C_7^{sb\ell\ell} - 8 \, \C_9^{sb\ell\ell} 
  + \frac{1}{2} \, \C_{7'}^{sb\ell\ell} - 8 \,\C_{9'}^{sb\ell\ell}\ , && \notag \\[2mm]
\frac{\alpha_{\rm em}}{4\pi} V_{tb}V^\star_{ts}\ C_{T5}^\ell &= \frac{1}{2} \,\C_7^{sb\ell\ell} - 8\, \C_9^{sb\ell\ell} 
  - \frac{1}{2} \, \C_7^{sb\ell\ell} + 8 \, \C_{9'}^{sb\ell\ell}\ . &&
\end{align}

\section{Comparison with the Literature}
\label{changebasis}

In this Appendix we compare our results from Section~\ref{ADM} with previously published results for the anomalous dimension matrices.

\subsection*{General Remarks}

Historically, the $\Delta F=1$ effective Hamiltonian does not contain all the operators in Eqs.~\eqref{eqn:fopq} and \eqref{eqn:fops}, but only the subset that corresponds to the low energy effective theory of the weak interactions in the SM. They are usually divided in three classes depending on the leading order mechanism that induces them in the full theory:\footnote{We ignore here for simplicity possible global normalization factors, as for example $V_{tb} V_{ts}^*$, since they do not affect the ADM.}
\begin{itemize}
  \item \emph{Current-current operators} arising from a tree-level exchange of a $W$-boson; we can denote them by
(in the notation of Eq.~({\ref{eqn:fopq}}))
    \begin{equation}
      \mathcal{P}_1^{u,c} = \fop_1^{u,c}, \qquad
      \mathcal{P}_2^{u,c} = \fop_2^{u,c}. 
    \end{equation}
  \item \emph{QCD-penguin operators} coming from penguin diagrams with a gluon exchange; 
    since the quark-gluon couplings are flavour independent, the summation over all possible quark flavours is taken:
    \begin{align}
      \mathcal{P}_3 &= \sum_{q}\fop^q_1, &
      \mathcal{P}_4 &= \sum_{q}\fop^q_2, \notag \\
      \mathcal{P}_5 &= \sum_{q}\fop^q_3, &
      \mathcal{P}_6 &= \sum_{q}\fop^q_4.
      \label{eqn:QCDpenguin}
    \end{align}
  \item \emph{EW-penguin operators} originating in the SM from photon or $Z$ penguins and box diagrams; they correspond to the combinations
    \begin{align}
      \mathcal{P}_7 &= \frac{3}{2} \sum_{q}e_q\fop^q_3, &
      \mathcal{P}_8 &=\frac{3}{2}  \sum_{q}e_q\fop^q_4, \notag\\
      \mathcal{P}_9 &=\frac{3}{2}  \sum_{q}e_q\fop^q_1, &
      \mathcal{P}_{10} &=\frac{3}{2}  \sum_{q}e_q\fop^q_2.
      \label{eqn:EWpenguin}
    \end{align}
\end{itemize}
The operators $\fop_{5,\dots,10}^q$ usually are not considered in the SM.

We will now compare the ADM matrices presented in Section~\ref{ADM} with the previously published results. 
As already stated in section~\ref{set-up}, in QCD the operators $\O^{sbqq}_i$ with $q=u,d,c$ mix through vertex-correction or closed penguin diagrams (see Figs.~\ref{fig:QCDvertexcorrection}-\ref{fig:QCDclosedpenguin}), while for $q=s,b$ they mix in addition with open penguins (see in Fig.~\ref{fig:QCDopenpenguin}).
However, the one-loop ADM of the operators~\eqref{eqn:QCDpenguin} and~\eqref{eqn:EWpenguin} receives contributions from both open and closed penguins since for $q=b,s$ also the operators with even indices participate. Moreover, penguin diagrams appear with a multiplicity factor $f$ given that more than one quark flavour is allowed in the loop.
For a comparison with previous published results, it is necessary therefore to extract these three contributions;
in some cases, when this is not possible, a comparison is performed by recombining our results in Section~\ref{ADM}
in order to obtain the ADM for the operators~\eqref{eqn:QCDpenguin} and~\eqref{eqn:EWpenguin}.
Usually the ADMs are expressed in the Fierz basis, $\hat \gamma_{\mysmall \fop}$, and must be converted into our basis by means of the transformation~\eqref{eqn:fierztonew}:
\begin{equation}
  \hat \gamma_{\mysmall \O} = \hat R^{-1} \hat \gamma_{\mysmall \fop} \hat R\ .
\end{equation}
Also for $q=s,b$ the ADM has to be reduced to the minimal basis by applying the Fierz identities in~\eqref{eqn:fierzids}
and by eliminating from the ADM the rows and/or the columns corresponding to the even (redundant) operators.
In the following it is understood that such transformations have to be applied in the comparison whenever necessary.

\subsection*{QCD mixing}

Early calculations of genuine vertex corrections to four-quark operators $\fop_{1-4}^q$ can be found in~\cite{Altarelli:1974exa,Gaillard:1974nj}. 
One- and two-loop ADM in QCD for $|\Delta S| =1$ were calculated in refs.~\cite{9211304,Ciuchini:1993vr}; the contribution of vertex correction diagrams to the mixing of $\fop_{1-4}^q$ can be read, for example, from the ADM of QCD and EW penguin operators~\eqref{eqn:QCDpenguin} in Section 3.1 of~\cite{9211304}.

In Ref.~\cite{Ciuchini:1997bw} the one- and two-loop ADM for $|\Delta F|=2$ were calculated; the results are expressed in terms of four-quark operators with a generic flavour structure, denoted by $Q_{1-5}^\pm$.
The vertex corrections for the operator $\fop_{5-10}^q$ can be extracted  by identifying the operators defined in Eq.~(13) of~\cite{Ciuchini:1997bw} with
\begin{align}
  Q_1^\pm &= \frac{1}{2}\fop_1^q \pm \frac{1}{2}\fop_2^q, &%
  Q_2^\pm &=  \frac{1}{2}\fop^q_3 \mp \fop_8^q, \notag \\
  Q_3^\pm &= \mp \frac{1}{4} \fop_4^q+\frac{1}{2} \fop_7^q, &%
  Q_4^\pm &= \frac{1}{2}\fop_5^q \mp \frac{1}{4} \fop_6^q \mp \frac{1}{16} \fop_{10}^q, \notag \\
  Q_5^\pm &= \pm 3 \fop_6^q -\frac{1}{2} \fop_9^q \mp \frac{1}{4} \fop_{10}^q;
\end{align}
the above relations take into account also that in~\cite{Ciuchini:1997bw} $\sigma^{\mu\nu}$ is defined as $\sigma^{\mu\nu} = \frac{1}{2} [\gamma^\mu,\gamma^\nu]$. When $q=b,s$ the $Q^-$ operators vanish and $Q^+_1=\fop_1^q, Q^+_2=\fop^q_3, Q^+_3=\fop^q_7, Q^+_4=\fop^q_5, Q^+_5=-\fop^q_9$.

The contribution to the $|\Delta F|=1$ ADM from one loop penguin where first evaluated
in~\cite{Shifman:1976ge,Gilman:1979bc,Guberina:1979ix}.
The penguin contributions to the Class-V ADM due to insertions of four-quark operators can be retrieved, for example,
from section 3.2 of~\cite{9211304}. The operators $\mathcal{P}_4$ and $\mathcal{P}_6$ can mix only via closed penguin
diagrams, so that the relative contribution to the ADM originating from the insertion of $\fop_{2,4}^q$,
with $ q=u,d,c$, is obtained by setting the number of flavours $f=1$ in the results for $\mathcal{P}_4$ and
$\mathcal{P}_6$. 
On the contrary, $\mathcal{P}_3$ can mix only through an open penguin; the ADM contribution due to~$\fop_1^{s,b}$ is
just one half of the result for $\mathcal{P}_3$.
Moreover, we note that the mixing of $\fop_7^{b,s}$ into $\fop_2^{q}$ via an open penguin is related through Fierz
identities~\eqref{eqn:fierzids} to the mixing pattern of the operators~$\fop_4^q$, from which the contribution to the
ADM can be extracted as well. The ADM of the operators $\fop^q_{5-10}$ were also calculated
in~\cite{Borzumati:1999qt,Buras:2000if}.

Vertex corrections to four-quark operators do not depend on the flavour, so they can be employed to calculate
directly the ADM of the operators in Classes I, III and IV. They also contribute to the diagonal sub-blocks $\hat A, \hat B$ and $\hat C$ of Class V  where, however, penguin contribution must be included as well: closed penguins for $\hat A^q$ and $\hat B^q$ and open penguins for $\hat C$. The off-diagonal sub-blocks of the Class-V ADM in~\eqref{eqn:ADMclassV} are generated only by penguins: closed penguins for the sub-blocks $\hat Z^q$ and $\hat H^q$ and open penguins for $\hat I^q$ and $\hat D$.

The one-loop QCD mixing of the operators $\O_{7\gamma}$ and $\O_{8g}$ appearing in the sub-block $\hat E$ in~\eqref{eqn:ADMclassV} was calculated in~\cite{Shifman:1976de,Grinstein:1990tj}. 
Given the normalization of the four-quark operators in Class~V, the only operators mixing into $\O_{7\gamma}$ and $\O_{8g}$ at $O(\alpha_s)$ are $\O^{sbcc}_{7-10}$ and $\O^{sbbb}_{5,7,9}$, corresponding to the sub-blocks $\hat K$ and $\hat J$ in~\eqref{eqn:ADMclassV}; the mixing was calculated in~\cite{Borzumati:1999qt}. 
We recall that in the SM, where only QCD and EW penguin operators are considered, the mixing between
$\mathcal{P}_3,\dots,\mathcal{P}_6$ and $\O_{7\gamma},\O_{8g}$ vanishes at one-loop. Therefore the leading
$O(\alpha_s)$ contribution to the ADM arises from two-loop diagrams, calculated
in~\cite{Ciuchini:1993fk,Ciuchini:1993ks}; with our conventions, these mixing contributions enter in the ADM only
at order~$\alpha_s^2$.

The QCD mixing of semileptonic operators in Classes II, V and VI (the
blocks $\hat \gamma_{\sss \rm II}$, $\hat{F}$ and $\hat \gamma_{{\sss \rm VI}a,b,c}$) is determined simply by the anomalous dimension of the quark current. 
The operators with vector current do not have an anomalous dimension in QCD due to current conservation. 
The ADM of scalar and tensor currents can be recovered from the results of Ref.~\citep{Gracey:2000am}.
The ADM of baryon-number violating operators in Class~VII are new to our knowledge;
a calculation with a UV cut-off can be found in Ref.~\cite{Buras:1977yy}.

\subsection*{QED mixing}

Electromagnetic corrections to the mixing of four-quark operators in the $|\Delta S|=1$ Hamiltonian were computed
in~\cite{Lusignoli:1988fz} at one-loop and in~\cite{Ciuchini:1993vr} at two-loops.
From Appendix A of~\cite{Lusignoli:1988fz} it is possible, for example, to extract the contributions to the ADM due to vertex corrections and the penguin diagrams. 
Vertex corrections are recovered from the ADM sub-block relative to the mixing of EW penguin operators into QCD penguin operators; it easy to see that such mixing is driven only by vertex corrections.
The sub-block of the ADM giving the mixing between QCD penguin operators and the EW ones yields the contributions
arising from closed penguin diagrams, which are denoted by the number of up- and down-type quarks $f_u$ and $f_d$,
and open penguins, given by the remaining $f_u, f_d$-independent part once the vertex corrections are subtracted.

Vertex corrections determine for the ADM entries of the operator $\mathcal{O}^{sbsb}_{1,4,5,1'}$ in Class~I, the sub-block $\hat \Gamma_\mathrm{III}^{1-4}$ of Class~III, and the entries relative to the operators $\O^{sbsd}_{1,3}$ in Class~IV. 
In Class~V, the sub-blocks $\hat A, \hat B$ ($\hat C$) receive a contribution from vertex corrections and closed (open and closed) penguins. The off-diagonal sub-blocks in~\eqref{eqn:ADMclassV} are generated by closed penguins for the sub-blocks $\hat Z^q$ and $\hat H^q$ and both open and closed penguins for $\hat I^q$ and $\hat D$.
Also the sub-block $\hat N$ and $\hat P$, giving the mixing of four-quark operators into semileptonic ones, can be obtained from $\hat H$ and $\hat I$ by appropriate substitution of the quark charge with the lepton charge.
In a similar way the results for $\hat G, \hat L$ and $\hat O$ can be derived from $\hat H$ by removing a factor of
three (the lepton in the loop does not carry color) and substituting the quark charges with the leptonic one.
The QED mixing of the magnetic operators $\O_{7\gamma}, \O_{8g}$,  the sub-block $\hat E$ of~\eqref{eqn:ADMclassV},
was calculated in~\cite{Baranowski:1999tq}, at one loop, and in~\cite{Bobeth:2003at} at two loops, where the mixing
of the semileptonic operators 
\begin{equation}
  Q_9 = \sum_\ell \O^{sb\ell \ell}_1
  \qquad \mbox{and} \qquad
  Q_{10} = \sum_\ell \frac{5}{3} \O^{sb \ell \ell}_1-\frac{1}{6} \O^{sb \ell \ell}_3
\end{equation}
is also presented. 
The QED mixing of semileptonic operators in Class V (corresponding to the blocks $\hat F$ and $\hat \gamma_{\rm Vb}$) can be recovered from the results of Ref.~\cite{Crivellin:2017rmk}, where the one-loop ADM of an effective Lagrangian for $\mu \to e$ transitions in calculated.

To our knowledge, the $O(\alpha_\mathrm{em})$ term in the ADM of Class~I (only for the operators $\O^{sbsb}_{2,3,2',3'}$), Class~II, the sub-block $\hat \Gamma_\mathrm{III}^{5-10}$ of Class~III  and Class~VII are new.
In Class~V, the results of the sub-blocks $\hat B$ and $\hat M$, and the entries of $\hat C$ relative to the operators $\O^{sbqq}_{5-9}$ are also new.


\end{document}